CDECK  ID>, INFORM.

                           H E R W I G

            a Monte Carlo event generator for simulating
        +---------------------------------------------------+
        | Hadron Emission Reactions With Interfering Gluons |
        +---------------------------------------------------+
      G. Marchesini, Dipartimento di Fisica, Universita di Milano
          I.G. Knowles(*), M.H. Seymour(+) and  B.R. Webber,
                   Cavendish Laboratory, Cambridge
------------------------------------------------------------------------
 with Deep Inelastic Scattering and Heavy Flavour Electroproduction by
 G.Abbiendi and L.Stanco, Dipartimento di Fisica, Universita di Padova
------------------------------------------------------------------------
        and Jet Photoproduction in Lepton-Hadron Collisions
             by J. Chyla, Institute of Physics, Prague
------------------------------------------------------------------------
(*)present address: Dept. of Physics & Astronomy, University of Glasgow
------------------------------------------------------------------------
(+)present address: Theory Division, CERN
------------------------------------------------------------------------
                    Version 5.9 - 22nd July 1996
------------------------------------------------------------------------
 Main reference:
    G.Marchesini,  B.R.Webber,  G.Abbiendi,  I.G.Knowles,  M.H.Seymour,
    and L.Stanco, Computer Physics Communications 67 (1992) 465.
------------------------------------------------------------------------
 Please send e-mail about  this program  to one of the  authors at the
 following addresses:
       Decnet   : 19616::webber, vxdesy::abbiendi, 19800::knowles
       Internet : webber@hep.phy.cam.ac.uk,
                  seymour@surya11.cern.ch, knowles@v2.ph.gla.ac.uk
------------------------------------------------------------------------

                         ****** CONTENTS ******

    1. INTRODUCTION
    2. NEW FEATURES OF THIS VERSION
    3. FEATURES NOT YET INCLUDED
    4. PROGRAM STRUCTURE
    5. BEAMS AND PROCESSES
    6. INPUT PARAMETERS
    7. COMMON BLOCK FILE
    8. FORM FACTOR FILE
    9. EVENT DATA
   10. STATUS CODES
   11. EVENT WEIGHTS
   12. HEAVY FLAVOUR DECAYS
   13. SPACE-TIME STRUCTURE OF EVENTS
   14. COLOUR REARRANGEMENT MODEL
   15. QCD HARD SUBPROCESSES
   16. DIRECT PHOTON SUBPROCESSES
   17. QCD HIGGS PLUS JET SUBPROCESSES
   18. ELECTROWEAK SUBPROCESSES
   19. INCLUDING NEW SUBPROCESSES
   20. ERROR CONDITIONS
   21. SAMPLE OUTPUT
   22. GUIDE TO SAMPLE OUTPUT

------------------------------------------------------------------------

                   ****** 1. INTRODUCTION ******

    HERWIG is a general-purpose event generator for high energy hadronic
    processes,  with particular  emphasis on the detailed  simulation of
    QCD parton showers.  The program has the following special features:

  * Simulation  of any  combination of  hard lepton,  hadron  or  photon
    scattering and soft hadron-hadron collisions in one package.

  * Colour coherence of partons (initial and final) in hard subprocesses

  * Heavy flavour hadron production and decay with QCD coherence effects

  * QCD jet evolution  with soft gluon interference via angular ordering

  * Backward  evolution of  initial-state partons including interference

  * Azimuthal correlations  within and  between jets due to interference

  * Azimuthal correlations  within  jets  due  to   gluon   polarization

  * Cluster  hadronization of jets via  non-perturbative gluon splitting

  * A complete space-time picture from parton showers to hadronic decays

  * A colour rearrangement model based on an events space-time structure

  * A similar  cluster  model for soft and  underlying  hadronic  events

   Further details may be found in the references cited above and at the
   end of this section, and in comments distributed throughout the code.

   The program  operates by  setting up  parameters in common blocks and
   then  calling a  sequence of  subroutines to generate an event. Para-
   meters  not set in the main program  HWIGPR are set to default values
   in the main initialisation routine HWIGIN.

   To generate events the user must first set up the beam particle names
   PART1, PART2 (type CHARACTER*8) in the common block /HWBEAM/, and the
   beam momenta  PBEAM1, PBEAM2 (in GeV/c), a process code IPROC and the
   number of events required MAXEV in /HWPROC/.  See section 5 for beams
   and processes available.

   All  analysis of  generated  events  (histogramming, etc.)  should be
   performed  by  the  user-provided  routines  HWABEG  (to initialise),
   HWANAL (to analyse an event) and  HWAEND  (to terminate).  At present
   HWANAL  writes  event  information and stable  particle  data on unit
   LWEVT  defined in  HWIGIN (or simply returns if LWEVT=0).  See HWANAL
   for details of event information written.  Note  that  HWANAL  should
   always begin with the line
      IF (IERROR.NE.0) RETURN
   to prevent it being executed for incomplete events.

   A detailed  event  summary is printed out for the first  MAXPR events
   (default MAXPR=1). Set  IPRINT=2  to list the particle identity codes
   and  (simplified) particle decay schemes used in the program.

   The  programming language is  standard Fortran 77 as far as possible.
   However,  the  following  may  require  modification  for  running on
   computers  other than  Vax's:

  *  Most  common  blocks are  inserted  by  INCLUDE 'HERWIG59.INC'  Vax
     Fortran  statements   (see  below  for  contents  of  HERWIG59.INC)

  *  Subroutine  HWUTIM (returning CPU time left) is  machine dependent.

    The principal references are:

    G.Marchesini  and  B.R.Webber,  Nucl. Phys. B310 (1988) 461;  I.G.
    Knowles,  Nucl. Phys. B310 (1988) 571;  S.Catani, G.Marchesini and
    B.R.Webber, Nucl. Phys. B349 (1991) 635;  G.Abbiendi and L.Stanco,
    Comp.Phys.Comm. 66 (1991) 16, Zeit. Phys. C51 (1991) 81;
    M.H.Seymour, Zeit. Phys. C56 (1992) 161.

    Some additional relevant references are:

    A.Bassetto, M.Ciafaloni and G.Marchesini, Phys. Rep. 100 (1983) 201;
    G. Marchesini  and  B.R. Webber,  Nucl. Phys.  B238 (1984) 1;  Phys.
    Rev. D38 (1988) 3419;  B. R. Webber,   Nucl. Phys.  B238 (1984) 492;
    Ann. Rev. Nucl. Part. Sci. 36 (1986) 253;  I.G. Knowles, Nucl. Phys.
    B304 (1988) 794; Computer Phys. Comm. 58 (1990) 271.
------------------------------------------------------------------------

            ****** 2. NEW FEATURES OF THIS VERSION ******

  * The common block file HERWIG59.INC has been significantly rearranged
    and tidied up.

  * Many new hadrons have been added.  All S & P wave mesons are present
    including the 1^P_0 & 3^P_1 states and many new, excited B^**, B_c &
    quarkonium states. Also all D wave kaons and some `light' I=3 states
    [pi_2, rho(1700) & rho_3].  All the baryons (singlet/octet/decuplet)
    containing up to one heavy (c,b) quark are included.

       --- Consequently the default parameters require retuning ---

  * New 8-character particle names have been introduced and  the revised
    7 digit PDG numbering scheme,  as advocated in the LEP2 report,  has
    been adopted.

  * The layout of HWUDAT  has been altered to make it easier to identify
    and modify particle propeties. Three new arrays have been introduced
    RLTIM, RSPIN & IFLAV. These are: the particle's lifetime (ps), spin,
    and a code which specifies the flavour content of each hadron - used
    (in HWURES) to create sets of iso-flavour hadrons for cluster decay.
    Using the standard numbering of quark flavours the convention is:

       mesons:          n_q n_qbar    Eg. pi^+:  21, pi^-:      12
       baryons: +/-n_q1 n_q2 n_q3     Eg. Xi^0: 332, Xi^0bar: -332  etc.
               (-ve for antibaryons; digits in decreasing order)

    Light, neutral mesons are identified as: 11 if I=1: pi^0,rho^0,...
                                             33 if I=0: eta, eta'.. etc.

    Some parts of the program have been automated so that it is possible
    for the user to add new particles by specifying their properties via
    the arrays in /HWPROP/ & /HWUNAM/ and increasing NRES appropriately:
    this should be done before a call to HWUINC.

    As an example following lines add an isoscalar, spin pi state 'STAN'
    and a (very light) stable toponium state 'BEER' with the decay mode:
    STAN ---> BEER+BEER+BEER.

      NRES=NRES+1
      RNAME(NRES)='STAN    '
      IDPDG(NRES)=666
      IFLAV(NRES)=11
      ICHRG(NRES)=0.
      RMASS(NRES)=0.5
      RLTIM(NRES)=1.000D-10
      RSPIN(NRES)=3.142
      NRES=NRES+1
      RNAME(NRES)='BEER    '
      IDPDG(NRES)=66
      IFLAV(NRES)=66
      ICHRG(NRES)=0.
      RMASS(NRES)=0.1
      RLTIM(NRES)=1.000D+30
      RSPIN(NRES)=0.0
      CALL HWMODK(666,1.D0,0,66,66,66,0,0)

  * The mixing angles of all the light, I=0 mesons can now be set using:

     ETAMIX:  eta       <-> eta'        F0MIX:  f_0(1300) <-> f_0(980)
     PHIMIX:  omega     <-> phi,        F1MIX:  f_1(1285) <-> f_1(1510)
     H1MIX:   h_1(1170) <-> h_1(1380)   F2MIX:  f_2       <-> f_2'

  * Using the logical arrays VTOCDK & VTORDK the production of specified
    particles can be stopped in both cluster decays and via the decay of
    other unstable resonances.

  * A priori weights for the relative production rates in cluster decays
    of mesons and baryons differing only via their S & L quantum numbers
    can be supplied using SNGWT & DECWT for singlet (i.e. Lambda-like) &
    decuplet baryons and REPWT for mesons. The old VECWT now corresponds
    to REPWT(0,1,0) and TENWT to REPWT(0,2,0).

  * The default masses of the c and b quarks have been lowered to 1.55 &
    4.95 repectively: this corresponds to the mass of the lightest meson
    minus the u/d quark mass. This increases the number of heavy mesons,
    and hence total multiplicities,  and slightly softens their momentum
    spectrum.  The rate of photoproduced charm states increases and B-pi
    momentum correlations become smoother.

  * The resonance decay tables supplied in the program have been largely
    revised. Measured/expected modes with branching fraction at or above
    1 per mille are given, including 4 & 5 body decays. To print the new
    tables call HWUDPR.

  * The arrays FBTM, FTOP & FHVY which stored the branching fractions of
    the bottom, top & heavier quarks' `partonic' decays are now nolonger
    used.  Such decays are  specified in the same way as all other decay
    modes: this permits different decays to be given to individual heavy
    hadrons.  Partonic decays of charm hadrons and quarkonium states are
    also now supported.  The products' order in a partonic decay mode is
    significant. For example if the decay is: Q --> W+q --> (f+fbar')+q,
    occuring inside a Q-sbar hadron the required ordering is:

             Q+sbar --->(f+fbar')+(q+sbar)
                     or (q+fbar')+(f+sbar)  `colour rearranged'

    In both cases the (V-A)^2 ME^2 is proportional to: p_0.p_2 * p_1*p_3

  * The structure of the program has been altered so that secondary hard
    subrocess and subsequent fragmentation associated with each partonic
    heavy hadron decay appear separately. Thus pre-hadronization t quark
    decays are treated individually  as are any subsequent bottom hadron
    partonic decays.

  * Additionally decays of heavy hadrons to exclusive non-partonic final
    states are supported. No check against double counting from partonic
    modes is included. However this isn't expected to be a major problem
    for the semi-leptonic and 2-body hadronic modes supplied.

  * An array NME has been introduced to enable a possible matrix element
    to be specified for each decay mode.

    NME = 0   Isotropic decay
          100 Free particle (V-A)*(V-A): p_0.p_2 * p_1.p_3
          101 Bound quark (V-A)*(V-A):   p_0.p_2 * p_1*[p_3 - xs*p_0]
                        xs = m_Q/M_0 - spectator quark momentum fraction
          130 Ore & Powell ortho-positronium ME^2: onium --> gg+g/gamma.

    The list of matrix elements presently supported is modest, users are
    urged to contact an author to have other MEs implimentated.

  * The decay tables can be written to/read from a file by using HWIODK,
    adopting the format advocated in the LEP2 report. In addition to the
    PDG numbering of particles the HERWIG numbers or character names can
    be used. This permits easy alteration of the decay tables. In HWUINC
    a call is made to HWUDKS which sets up HERWIGs internal pointers and
    performs some basic checks of the decay tables. Each decay mode must
    conserve charge and be kinematically allowed  and not contain vetoed
    decay products. The sum of a particles branching ratios is set to 1.
    Also  a warning is printed  if an antiparticle does not have all the
    charge conjugate decays modes of the particle.

  * HWMODK enables changes to the decay tables to be made by alterating/
    adding single decay modes including on an event by event basis. This
    can be done before HWUINC, in which case when altering the BR and/or
    ME code of an existing mode a warning is given of a duplicate second
    mode which supercedes the first. BRs set below 10^-6 are eliminated,
    whilst if one mode is within 10^-6 of 1 all other modes are removed.
    Note  that some forethought is required if the BRs of 2 modes of the
    same particle are changed since  the operation of rescaling to 1 the
    BR sum causes a non-commutativity in the order of the calls.

  * Production vertex information is now made available, using VHEP, for
    all partons, clusters and final state particles: set PRVTX=.TRUE. to
    print them. The vertices of partons and clusters are given wrt local
    coordinates associated with their individual hard sub-process.

  * All partonic and resonace rest frame lifetimes are generated with an
    exponential distribution: exp(-t/<tau>)/<tau>. The average lifetime,
    <tau>, is given in terms of the particles mass, width and virtuality
    by:
                              hbar.sqrt(q^2)
        <tau>(q^2) =   ----------------------------- 
                     \/(q^2-M^2)^2 + (Gamma.q^2/M)^2

                   = hbar/Gamma        for an on-shell       particle
                   ~ hbar.q/(q^2-M^2)      a  highly virtual particle

    For partons an effective width = sqrt(VMIN2), to act as a cut-off on
    lifetimes, is introduced.

  * The space-time picture for cluster formation and splitting is partly
    ad hoc and partly string inspired - no physics depends upon it.

  * All particles with lifetimes greater than PLTCUT are set stable.

  * If PIPSMR=.TRUE. the primary interaction point's spatial position is
    is smeared according to the triple Gaussian in HWRPIP: this position
    is assigned to the CMF track.

  * If MAXDKL=.TRUE. then each putative decay is tested in HWDXLM to see
    that it occurs within a specified volume (cylinder/sphere for IOPDKL
    =1/2): if not it is set stable.

  * If MIXING=.TRUE. then  B^0_d,s mesons are allowed to oscillate: XMIX
    and YMIX contain Delta-M/Gamma and Delta-gamma/2*Gamma respectively.
    A new particle, ISTHEP=200,  is introduced giving the flavour of the
    neutral B meson at production in addition to the `decaying' track.

  * A multiple intra & inter-jet colour rearrangement model is available
    for  CLRECO=.TRUE.  The q-qbar pairings in two non-adjacent clusters
    are interchanged with probability PRECO if the distances between the
    production vertices of both q-qbar pairs when added in quadrature is
    reduced. EXAG can be used to artificially scale the lifetimes of any
    weak bosons.

  * A number of bugs have been corrected: in HWEPRO for weighted events;
    in  HWSBRN  affecting the reconstruction of the photon beam remnant;
    and in HWHEPG stopping event generation. Plus minor modifications to
    HWBGEN;  in the use of HWHIGM by HWHIGJ; and small changes in HWHDIS
    & HWHEGG.

  * A significant bug in HWDHQK, affecting top quark decays, was present
    in version 5.8 ONLY.  The scale of the top decay had been set to the
    b-quark mass,  stopping gluon radiation  from the b  and restricting
    that from the W decay products to have transverse momentum less than
    the b mass. The scales are now correctly set for top decays.

  * Improved efficiency of photon generation in HWEGAM.

  * New hard sub-process have been added:

    - Compton scattering, gamma + q --> gamma + q, IPROC=5300.

    - Two-to-two  parton  scattering  via  exchange  of a colour singlet
      IPROC=2400 Mueller-Tang pomeron: the fixed alpha_s and omega_0 are
      given by ASFIXD and OMEGA0 respectively.
      IPROC=2450 photon exchange, for like flavour qqbar pairs including
      the t-channel component of the interference with q-qbar -> q-qbar.

    - Drell-Yan has been extended to the production of all fermion pairs
      IPROC=1399; 1300 gives all quark flavours 1300+IQ a specific quark
      flavour, 1350 all leptons (including neutrinos) 1350+IL a specific
      lepton flavour.   The s-channel component of the interference with
      like flavour q-qbar scattering is included here.

    - Z+jet production is included as IPROC=2150 (HWHW1J becomes HWHV1J)

  * Running coupling now used for prompt J/PSI production in DIS.

  * The phase-space limits for the momentum fraction of incoming photons
    in the Weizsacker-Williams approximation is now set by the variables
    YWWMIN  &  YWWMAX,  allowing  different  ranges  for  the tagged and
    untagged photons in two-photon DIS.

  * Interfaced to the  Schuler-Sjostrand  parton distribution functions,
    version 2.  These appear as  PDFLIB  sets with author group 'SaSph',
    but are actually implemented via a call to their  SASGAM  code.  The
    value in MODPDF specifices the set (1-4 for 1D [recommended set],1M,
    2D,2M), whether the Bethe-Heitler process is used for heavy flavours
    (add 10), whether the P^2-dependence is included (add 20), and which
    of their P^2 models is used (add 100 times their IP2 parameter).

  * New variables ANOMSC(1 or 2,IBEAM) record the evolution scale and Pt
    at which an anomalous (gamma* --> q+qbar) splitting was generated in
    the backward evolution of beam IBEAM. set 0 if no such splitting was
    generated. This is implemented in HWBGEN and HWSBRN.

  * In preparation for multiple interactions, several routines have been
    added or modified.  New are:  HWHREM for identifying and cleaning up
    the beam remnants;  HWHSCT to administer the extra scatters.   Minor
    modifications to:  HWBGEN & HWSBRN, don't report energy conservation
    errors when ISLENT = -1;  HWSSPC, improved approximation for remnant
    mass at high energies;  and HWUPCM, improved safety against negative
    square roots.

  * Photon Initial State Radiation in  e+e- annihilation events allowed.
    TMNISR sets the minimum s-hat/s value,   ZMXISR sets the (arbitrary)
    separation between unresolved and resolved emission;  using ZMXISR=0
    switches off photon ISR.

  * Numerical integral in HWBDED now done analytically removing the need
    to reintegrate for each new energy; in principle allowing use in  5-
    jet WW events, but this is not yet implemented.

  * New phase-space variable WHMIN added.  This sets the minimum allowed
    hadronic mass and affects photoproduction reactions  (gamma-hadron &
    gamma-gamma) and DIS.  In lepton-hadron DIS it is largely irrelevant
    since there is already a cut on Bjorken y which at fixed s is almost
    the same but for lepton-gamma DIS it makes a big difference.

  * A new treatment of running Higgs width and non-resonant diagrams, as
    suggested in M.H. Seymour, Phys. Lett. B354 (1995) 409.  Selected by
    setting IOPHIG=2 or 3 (default);  previous options 2 and 3 have been
    withdrawn. Note that including the non-resonant diagrams changes the
    meaning of what is generated: IOPHIG = 0  or  1, gives the s-channel
    diagram,  an unphysical choice of part of the amplitude;  IOPHIG = 2
    or 3,  gives the I=0 & J=0 part of the excess over the cross section
    expected for a zero mass Higgs boson,  a physical choice  of part of
    the cross section. The inclusion of non-resonant diagrams causes the
    cross section to increase below and decrease above resonance.

  * New treatment of the splitting in two of clusters  containing hadron
    (or photon) remnants.  Previous versions gave the 2 fragments a mass
    spectrum typical of soft processes: dn/dm**2 = Gaussian.  In the new
    version  the child  containing the remnant  is treated as before but
    the other cluster, containing a perturbative parton, is treated as a
    normal clusters: dn/dm = m**psplt. IOPREM controls this behaviour: 0
    = old version, 1 = new (default).

  * Direct gamma+gamma* -> q+qbar is included in the hard correction for
    lepton-gamma DIS; plus minor bug fixed in HWBDIS.

  * The dummy routine  IUCOMP  has been removed, this avoids errors when
    the program is linked to CERNLIB.

  * It  has  been  noticed  that differences in the way quark masses are
    treated  in  different  processes  can cause inconsistencies between
    different  ways of generating the same process.  The most noticeable
    example is in direct photoproduction, where one can use process 9130
    or  5000.  See the note at the end of Section 5 of the documentation
    for more information on the strategies used in different processes.

    Version 5.1 of  HERWIG was described  in detail in  Computer Physics
    Communications 67 (1992) 465. For completeness we list here also the
    main new features added in versions 5.2 - 5.7.

    In version 5.2:

  * New e+e- processes:
    - two photon processes, IPROC = 500+ID  where ID=0-10 is the same as
      in Higgs processes for qqbar, llbar, and W+W-.  The phase space is
      controlled  by  EMMIN,EMMAX for the CMF mass,  PTMIN,PTMAX for the
      transverse momentum of the CMF in the lab,  and CTMAX  for the CMF
      angle of the outgoing particles.
    - photon-W fusion,  IPROC = 550+ID  where  ID=0-9 is the  same as in
      Higgs processes,  except that ID=1 or 2 both give the sum of dubar
      and udbar etc.  The phase space is controlled by EMMIN,EMMAX only.
      The full 2-->3 matrix elements for photon e-->f f'bar nu are used,
      so the cross section  for real W production is correctly included.
    - ZZ pair production,  IPROC=250 is treated just like WW production,
      and is based on the program kindly supplied by Zoltan Kunszt.

  * New ep processes:
    - the phase space for BGF is now controlled by EMMIN,EMMAX as above.
      The default values are 0 and RootS respectively,  corresponding to
      the behaviour of version 5.1
    - J/psi production from BGF, IPROC = 9104 is now available.
    - W W fusion to Higgs is now available in ep, IPROC = 9500+ID.

  * IPROC = 1600+ID now gives the sum of gluon fusion and q qbar fusion.
    This is especially important in e+e- if tan(beta) is large,  when it
    is dominated by e+e- --> e+e- gamma gamma --> e+e- b bbar H.

  * Users can now force  Z --> b bbar  decays,  with MODBOS(i)=7  (for a
    complete list see section 18).  For example, IPROC=250, MODBOS(1)=7,
    MODBOS(2)=0 gives ZZ production with one Z decaying to b bbar.

  * All Higgs vertices now include an enhancement factor to  account for
    non-SM couplings. ENHANC(ID), where ID=1-11 is the same as for Higgs
    production, holds the ratio of the AMPLITUDE for the given vertex to
    that of the SM. This of course only simulates the chargeless scalars
    of any extended model, and not the pseudoscalars or charged Higgses.

  * The heavy quark  content of the photon  now uses the  corrections to
    the Drees-Grassie distribution functions for light quarks,  recently
    calculated by C.S.Kim et al. (see M.Drees & C.S.Kim, DESY 91-039 and
    C.S.Kim, Durham preprint DTP/91/16).

  * A new structure function set, Owens1.1, similar to Duke+Owens1, but
    fitted to new data (Preprint FSU-HEP-910606) is available via
    NSTRU=5, and is now the default structure function set.

    In version 5.3:

  * O(alpha-s) jet production in ep processes has been included (IPROC=
    9200 etc),  with Q**2 range controlled by  Q2MIN, Q2MAX and minimum
    jet transverse momentum  (in the hard subprocess c.m. frame) set by
    PTMIN. The new subroutines were written by Sebastian Brandis and we
    are grateful to him for permission to use his code.

  * Minor bugs have been fixed in the  backward evolution of quarks into
    photons,  hadronic  processes  in e+e-, remnant hadronization in ep,
    and in the  generation of weighted events  (ie. with NOWGT=.FALSE.).

    In version 5.4:

  * A correction to  hard gluon  emission in e+e- events  has been added
    and is now the  default  process.  This uses the  O(alpha-s)  matrix
    element to add events in the  `back-to-back'  region of  phase-space
    corresponding to a quark-antiquark pair  recoiling from a  very hard
    gluon.  Although this is asymptotically  negligible,  and  cannot be
    produced  within the shower itself,  it has a sizeable effect at LEP
    energies.   As a result,  the default parameters  have been retuned,
    and show a marked improvement in agreement with  OPAL data for event
    shapes sensitive to  three-jet  configurations  (J.W. Gary,  private
    communication).   The  uncorrected  process  has  been  retained for
    comparative purposes and is available as IPROC=120+IQ.

  * Photons are now included in  time-like parton showering.  The infra-
    red cutoff is VPCUT,  which defaults to SQRT(S)  corresponding to no
    emission.  Agreement with LEP data is satisfactory  if used together
    with the  matrix element correction  to produce photons in the back-
    to-back region.  The results are  insensitive  to  VPCUT  variations
    in the range 0.1-1.0 GeV.

  * W decay correlations and width are  now correctly included in  W+jet
    production (previous versions used unpolarized, on-shell approx.).

  * An inconsistency in  the argument used for  alpha_s in the branching
    g -> q qbar has been removed. The change is a non-leading correction
    but leads to slightly more quarks in gluon jets.

  * A new parameter  B1LIM  has been introduced for  B cluster hadroniz-
    ation.  If  MCL is the B cluster mass and  MTH the threshold for its
    decay into 2 hadrons,  the probability of  its decay into a single B
    hadron is:   1 if MCL<MTH,   0 if MCL>(1+B1LIM)*MTH,   with a linear
    interpolation i.e. 1-(MCL-MTH)/(B1LIM*MTH) if MTH<MCL<(1+B1LIM)*MTH.
    Thus the default value  B1LIM=0 gives the same as previous versions,
    while B1LIM>0 gives a harder B spectrum.

  * B decays can now be performed by the  EURODEC  or  CLEO  Monte Carlo
    packages.  The new variable  BDECAY  controls which package is used:
    'HERW' for HERWIG; 'EURO' for EURODEC; 'CLEO' for CLEO.  The EURODEC
    package can be obtained from the CERN library.   The CLEO package is
    available by  kind permission of the CLEO collaboration,  and can be
    obtained from Luca Stanco at the address given above.

    In version 5.5:

  * The Sudakov form factors can now be calculated using the one-loop or
    two-loop alpha_s, according to the variable SUDORD (DEFAULT=1).  The
    parton showering  still incorporates  the two-loop alpha_s in either
    case but if SUDORD=1 this is done using the veto algorithm,  whereas
    if SUDORD=2 no vetoes  are used in the  final-state evolution.  This
    means that the  relative weight of any  shower configuration  can be
    calculated in a closed form, and hence that showers can be `forced'.
    For example,  a package of  routines  should  be available  soon for
    forcing jets to contain  photons,  which will therefore  drastically
    improve the efficiency of photon FSR studies.
    To next-to-leading order  the two possibilities SUDORD=1 or 2 should
    be identical,  but they  differ  at beyond-NLO,  so some results may
    change a little. Previous versions were equivalent to SUDORD=1.

  * Alpha_em is now multiplied by the factor ALPFAC (DEFAULT=1)  for all
    quark-photon vertices in jets, and in the `dead zone' in e+e-.  This
    is a cheap  way of improving  the efficiency of  photon FSR studies,
    which should  not be needed  once photon forcing is available.  Note
    that results at small ycut become sensitive to ALPFAC above about 5.

  * A new parameter CLPOW (DEFAULT=2) is available in the cluster hadro-
    nization model.  A cluster of mass  MCL made of quarks of mass M1,M2
    is split into lighter clusters before decaying if
                 MCL**CLPOW > CLMAX**CLPOW + (M1+M2)**CLPOW
    Thus the previous value was CLPOW=2,  like the new default.  Smaller
    values will  increase the  yield of  heavier clusters  (and hence of
    baryons) for heavy quarks, without affecting light quarks much.  For
    example, the default value gives no b-baryons (for the default value
    of CLMAX) whereas CLPOW=1.0 makes b-baryons/b-hadrons about 1/4.

  * The event record has been modified to retain entries for all partons
    before hadronization  (with status ISTHEP=2).  During hadronization,
    the gluons are split  into quark-antiquark,  while other partons are
    copied to a location  (indicated by JDAHEP(1,*)) where their momenta
    may be shifted slightly, to conserve momentum,  during heavy cluster
    splitting. Previously the original momenta were shifted, so momentum
    appeared not to be conserved at the parton level.

  * Minor improvements have been made to: NLO correction to Higgs decays
    to qqbar;  pt spectra of outgoing electrons in two-photon processes;
    quark-mass effects in gamma-W fusion; WW spectrum below threshold in
    e+e-; t-bbar spectrum in W Drell-Yan (IPROC=1406).

  * Bugs preventing the use of  Sudakov form factor tables from disk and
    gluon-> diquarks splitting option under some circumstances, together
    with other minor bugs and machine-dependences, have been fixed.

    In version 5.6:

  * Decays of very heavy quarks  (top and higher generations)  can occur
    either before or after hadronization. At present all top quarks will
    decay  before/after hadronizing if the top mass is greater/less than
    130 GeV.  This can be changed in subroutine HWDTOP.  All higher (>3)
    generations now decay before hadronization. Note that the new state-
    ment  CALL HWDHQK  must appear in the main program between the calls
    to  HWBGEN and HWCFOR  to carry out any decays before hadronization.

  * Bugs in the subroutine  HWHDOA for O(alpha_s)  jet production in DIS
    have been corrected by J. Chyla,  who has also extended this process
    into the photoproduction region. If Q2MIN.LT.2D-6 (the new default),
    the kinematic lower limit on Q**2 is computed and used.  New options
    IPROC=9250 to 9277 use various approximations to the neutral-current
    matrix element, as specified in the Table below.

  * The photoproduction  processes  have  also  been  extended  from the
    original  heavy quark production program,  to include all quark pair
    production  (IPROC=9100-9106)  and QCD Compton (IPROC=9110-9122), as
    well as the sum of the two  (IPROC=9130).  The possible flavours for
    the 9100,9110 and 9130 processes are limited by the input parameters
    IFLMIN and IFLMAX  (defaults are 1 and 3, i.e. only u,d,s flavours).
    The corresponding Charged Current processes are now provided via the
    IPROC=9140-9144 codes.

  * All the DIS processes  IPROC=9000-9599  are now available in e+e- as
    well as  lepton-hadron  collisions.  The program  generates a photon
    from the second beam (only) in Weizsacker-Williams approximation and
    uses Drees-Grassie structure functions for DIS on the photon.

  * Pointlike photon-hadron scattering  to produce QCD jets is available
    as IPROC=5000.  This is suitable for  fixed-target  photoproduction,
    provided events  are generated in  a frame in  which the  target has
    high momentum, and then boosted back to the lab.  IPROC=5000+IQ gen-
    erates only those processes involving quark flavour IQ,  using exact
    kinematics and  light-cone  momentum fraction.  In both cases, after
    event generation the hard subprocess code  IHPRO is set to  51,52 or
    53 for photon+q->g+q, photon+qbar->g+qbar, or photon+g->q+qbar.

  * The default limits on Q**2 in DIS processes  (Q2MIN,Q2MAX) have been
    set very  small/large  (0.0, 1.D10)  and are reset  to the kinematic
    limits unless  changed by the user.  This means the default Q2MIN is
    not suitable for simple NC DIS (IPROC=9000 etc),  but is appropriate
    for jet and heavy quark photoproduction.

  * A new parameter  NMXJET,  the maximum number of  outgoing partons in
    a hard subprocess  (default 200)  has been  introduced in the common
    block file HERWIG56.INC.

  * For technical reasons, some HERWIG status codes  ISTHEP  between 153
    and 165 have changed their meanings. See the Table in sect.10 below.

  * Bugs in the hadronization of  diquark-antidiquark clusters have been
    fixed. Any such clusters with  masses below threshold for decay into
    baryon-antibaryon are shifted to the  threshold via a transfer of 4-
    momentum to a neighbouring cluster.

  * A bug in the default pion structure function (no gluons) is fixed.

    In version 5.7:

- ELECTRO-WEAK COUPLINGS: New arrays QFCH(16), VFCH(16,2), AFCH(16,2)
  and VCKM(3,3) have been set up for couplings and CKM matrix.  See the
  documentation file or HWIGIN for conventions.  Note that universality
  is not assumed, so lepton axial couplings may differ for example; this
  is primarily to cover Z' possibilities, see below. The variable
  SCABI=sin^2 theta_Cabibbo is however also retained for the present.

- A Z' has been introduced with PDG code 32, HERWIG identifier 202,
  default mass 500 GeV, width GAMZP (default 5 GeV) and name 'Z0PR'.
  It is invoked by setting ZPRIME=.TRUE. (default .FALSE.).

- POLARISATION: incoming lepton and antilepton beam polarisations
  are now specified by setting two new vectors EPOLN(3) and PPOLN(3):
  component 3 is longitudinal and 1,2 transverse. Transverse only occurs
  in e+e- routines; recall that two transverse 'measurements' are needed
  to see an effect so it should not arise elsewhere. Note that in DIS
  processes you have to set either EPOLN if it is a lepton or (exclusive)
  PPOLN if an antilepton.

  Polarisation effects are now included in e+e- 2/3 jet production
  and Bjorken process, together with DIS processes apart from J/psi
  production.

- NEW SUBPROCESSES:
  2200    QCD direct photon pair production (inc. g+g->gamma+gamma)
  5100+IQ Point-like photon/QCD heavy flavour pair production
  5200+IQ Point-like photon/QCD heavy flavour single excitation
  The latter two replace 5000+IQ, while 5000 remains as before (ie
  a sum over all processes and flavours with simplified kinematics)

- The kinematic reconstruction of DIS processes can now take place in
  the Breit frame, if BREIT=.TRUE. (the default value). Previous versions
  used the lab frame. Although the reconstruction is fully invariant under
  Lorentz boosts along the incoming hadron's direction, it is not under
  transverse boosts, so there should be some difference between the two
  frames. The boost is not performed for very small Q^2 (<10^-4) to avoid
  numerical instabilities, but the two frames are in any case equivalent
  for such small Q^2.

- A new parameter PRSOF to produce an underlying event in only a fraction
  PRSOF of events (default=1.0).  IPROC=19000 etc are thus equivalent to
  PRSOF=0.

- Non-diffractive hadronic minimum bias events (IPROC=8000) can now be
  generated for a wider variety of beams (P,PBAR,PI+/-,K+/-,E+/-,MU+/-,GAMA
  on target P; also P and PBAR or leptons on target N). The event weight
  (previously set to 1.0 for this process) is the estimated cross section
  based on the parametrizations of Donnachie and Landshoff, CERN-TH.6635/92.
  The non-diffractive cross section is assumed to be 70
  For lepton beams a photon is first generated using the effective photon
  approximation (see below) and then the on-shell photon cross section
  is used.

- A bug has been fixed in HWBRAN and HWSBRN (present in versions 5.1 to
  5.6) that led to too much transverse momentum being developed by the
  parton showers in hadron-hadron collisions. All radiation with pt
  greater than the hard process scale is now vetoed. In the case of
  initial-state radiation, this affects all events, while for final-state
  radiation it only affects those in which the two jets have a rapidity
  difference of more than about 3.4.

- When SUDORD=2, no veto is needed for gluon splitting to quarks. This
  means that no vetoes are needed for final state showering, except for
  the previously-mentioned transverse momentum cut. The removal of
  vetoes allows preselection of the flavours that a jet will contain,
  giving a huge increase in the efficiency of rare process simulation. A
  package is already available to simulate heavy flavour production
  inside jets, and the equivalent for photons should soon be available.

- Parameter BTCLM is now available to users to adjust the mass parameter
  in remnant formation. Its default value, 1.0, is identical to previous
  versions.

- There is a new switch CLDIR for cluster decays. CLDIR=0 is the same as
  previous versions, while CLDIR=1 (the default) means that a cluster that
  contains a `perturbative' quark, ie one coming from the perturbative
  stage of the event (the hard process or perturbative gluon splitting)
  `remembers' its direction: when the cluster decays, the hadron carrying
  its flavour continues in the same direction (in the cluster c.m. frame)
  as the quark. This considerably hardens the spectrum of heavy hadrons,
  particularly of c- and b-flavoured hadrons. It also introduces a tendency
  for baryon-antibaryon pairs preferentially to align themselves with the
  event axis (the `TPC/2gamma string effect').

- The functionality of the routine HWUINE has now been split between it
  and a new routine, HWUFNE. A call to the latter MUST be inserted into
  the users main program, between the calls to HWMEVT and HWANAL. A
  check is built in to version 5.7 to prevent execution if this change
  is not made. See the documentation file for an example main program.
  We should also take this opportunity to remind users that the analysis
  routine HWANAL should begin with the line
      IF (IERROR.NE.0) RETURN
  since if an event is cancelled, each of the routines is still called
  in turn until reaching the end of the main loop.

- If the new flag USECMF is .TRUE. (the default), events are boosted to
  their centre-of-mass frame before processing if necessary, and boosted
  back afterwards.  This second boost is performed by the new routine
  HWUFNE, so it is essential that this is inserted in the correct place,
  as described above.

- In hadronic processes with lepton beams (eg photoproduction in ep),
  the lepton->lepton+photon vertex now uses the full tranverse-momentum-
  dependent splitting function, with exact light-cone kinematics (i.e.
  the Equivalent Photon instead of the Weizsacker-Williams approximation).
  This means that the photon-hadron collision has a transverse momentum
  in the lepton-hadron frame, and must be boosted to a frame where it
  has no transverse momentum. Thus the cmf boost described above is
  always used in these processes, regardless of the value of USECMF.
  The correct lower energy cut-off appropriate to the hadronic process
  is applied to the photon, rather than the fixed cut of 5 GeV that
  was used in previous versions. The Q**2 of the photon is generated
  within the kinematically allowed limits, or the user-defined limits
  Q2WWMN and Q2WWMX (defaults 0 and 4) whichever is more restrictive.
  The momentum fraction is generated within the kinematic limits or
  between YBMIN and YBMAX (defaults 0 and 1).

- Point-like photon processes (IPROC=5***) are now also available with
  lepton beams, using the Equivalent Photon Approximation.

- Several minor improvements have been made to the O(as) processes in
  DIS (IPROC=91**):
  - A sign error has been corrected that led to the incorrect sign for
    the lepton-jet azimuthal correlation in QCD Compton processes.
  - An additional cut on the phase-space generation has been provided:
    the Bjorken-y variable (=Q^2/xs) is limited to range [YBMIN,YBMAX].
  - BGSHAT=.FALSE. is now the default.
  - J/Psi production (IPROC=9107) now uses the EPA instead of the WWA,
    with the same phase-space cuts as hadronic processes with lepton
    beams, see above.

- Many bugs have been fixed in the other O(as) process routines, HWHDOA
  and HWHDOM, ie for IPROC=92**. However, this process is no longer
  supported, and is only retained for comparative purposes. It will be
  withdrawn completely at the next version release.

- An interface is now provided to Mark Gibbs' HERBVI package for baryon-
  number violation, and other multi-W production processes, IPROC=7***.

- Minor bug fixes in HWHDIS, HWHEGW and HWHIGW and minor improvements in
  HWHHVY, HWHPHO, HWHQCD and HWHWEX hard process routines.

- New fictional e+e- processes: e+e- -> gluon+gluon(+gluon), IPROC=107
  & 127, treated just like e+e- -> quark+antiquark, summed over light
  quark flavours, for direct comparisons between quark and gluon jets.

- New logical variable PRNDEC (default=.TRUE. unless NMXHEP>9999) causes
  track numbers in event listings to be printed in hexadecimal if.FALSE.
  This is necessary for very large events such as those generated by the
  HERBVI package (see above).

- PDFLIB structure functions can now be used for the photon as well as
  nucleons. The new variable MODPHO acts just like MODPDF. PDFLIB calls
  have also been updated to allow for structure function sets with
  flavour-asymmetric sea contributions.

- A logical inconsistency has been fixed in the decays of clusters to
  eta or eta' - previously all mixing was neglected, leading to double-
  counting and a significant over-estimate of the number of each. The
  new variable ETAMIX gives the eta_8/eta_0 mixing angle in degrees
  (default = -20). Rates are not very sensitive to its exact value, as
  the eta'/eta suppression is dominated by mass effects in the cluster
  model.

- The maximum weight is now always printed in full precision (needed
  to be sure of generating the same events in repeated runs).

- New constants:  GEV2NB=389385 
                  ALPHEM(1)=1./137(.03599)    for Q^2=0.
                  ALPHEM(2)=1./128            for Q^2~M_W^2
  are introduced in various cross section formulae, and G_Fermi is
  eliminated.

- The default top quark mass was increased to 150 GeV.

    In version 5.8

  * A hard matrix element correction has been introduced in DIS (IPROC =
    90**).  This is switched  on and  off by the logical variable HARDME
    (default = .TRUE.).  The method is essentially identical to the e+e-
    correction,  generating  first  order  matrix-element  events  in  a
    phase-space region  complementary  to that of the parton shower. The
    e+e- correction is also now controlled by HARDME for consistency.

  * Soft matrix element corrections have been introduced in DIS and e+e-
    processes.  These correct  the distribution of emissions  within the
    parton  shower  phase-space.   It is similar to  the method  used in
    JETSET,  except that the  HARDEST emission is matched to the leading
    order matrix element, not the first as in JETSET.  This ensures that
    the correction  enters into the  form factor,  and not just the real
    emission probability.

  * In the backward evolution of initial-state radiation for photons the
    anomalous branching q-qbar <-- gamma has been introduced.

  * The treatment of forced branching of gluons and sea (anti-)quarks in
    backward evolution has been improved,   by allowing it to occur at a
    random scale  between the space-like cutoff  QSPAC  and the infrared
    cutoff, instead of exactly at QSPAC as before.
    A new option  ISPAC=2  allows the freezing of structure functions at
    the scale QSPAC,   while evolution continues to the infrared cutoff.
    The default,  ISPAC=0  is equivalent to previous versions,  in which
    perturbative evolution stops at QSPAC.

  * It is now possible to completely switch off initial-state radiation,
    by setting NOSPAC =.TRUE.   Only the forced splitting of non-valence
    partons is generated. The default is (of course) NOSPAC =.FALSE.

  * An option to damp the parton distributions of off mass-shell photons
    relative on-shell photons,  according to the scheme defined in Drees
    and Godbole MAD/PH/819 has been introduced. The adjustable parameter
    PHOMAS defines the crossover from the  non-suppressed to  suppressed
    regimes.  Recommended values lie in the range  QCDLAM to 1 GeV.  The
    default value PHOMAS=0. corresponds to no suppression as in previous
    versions.

  * The interface to PDFLIB version 4 has been slightly changed. Instead
    of indicating a PDF set by a unique number, an `author group' string
    and set number are required. PDFLIB version 3 can still be used from
    HERWIG, simply by setting the author group to 'MODE'. It is also now
    possible to independently set the PDF set for each of the two beams.
    For example, if you previously used MRS D- for the proton and Gordon
    -Storrow set 1 for the photon, by setting
          MODPDF=47
          MODPHO=231
    You should now set
          AUTPDF(2)='MRS'
          MODPDF(2)=28
          AUTPDF(1)='GS'
          MODPDF(1)=2
    Alternatively, if you are still using PDFLIB version 3, you can set
          AUTPDF(2)='MODE'
          MODPDF(2)=47
          AUTPDF(1)='MODE'
          MODPDF(1)=231

  * In the  CLDIR=1  option for  cluster  decays  a new parameter  CLSMR
    (default = 0.)  allows a  Gaussian smearing  of the direction of the
    perturbative quark's momentum.  The smearing is actually exponential
    in 1-cos(theta) with mean CLSMR.  Thus increasing CLSMR decorrelates
    the cluster decay from the initial quark direction.

  * New subprocess have been added:

    - The direct,  higher twist,  production of light (u,d,s) L=0 mesons
      by point-like photons is now available: IPROC = 5500 all Spin =0,1
      mesons, = 5510 only S=0 mesons; = 5520 only S=1 mesons. The vector
      mesons are produced with transverse  or  longitudinal polarisation

      and decayed accordingly.

    - High transverse momentum,  scalar Higgs production, in association
      with a jet, is now available as IPROC =2300. Only the top quark is
      included in the loops with IAPHIG controlling the approx. used: =0
      zero top mass limit; = 1 exact result; = 2 infinite top mass limit
      (default 1). Note the routines: HWHGJ1, HWHGJA, HWHGJB/C/D, HWUCI2
      and HWULI2 use (non-standard FORTRAN-77)  DOUBLE COMPLEX variables
      which may not be accepted by some compilers.   Users can change to
      COMPLEX variables, however this involves a risk of rounding errors
      spoiling numerical cancellations.

    - DIS with neutrino beams is now available in processes IPROC= 90**.

  * The DIS O(alpha_s) jet production processes, IPROC = 92**, have been
    withdrawn and are no longer supported.

  * A running electromagnetic coupling has been introduced,  HWUAEM(Q2).
    ALPHEM (now a single variable)  sets the Thomson limit (Q2=0) value,
    default = 0.0072993 (1/137.0).

  * Two new particles have been created: 'REMG', IDHW=71, IDHEP=9998 and
    'REMN',  IDHW=72,  IDHEP=9999   are  remnant  photons  and  nucleons
    respectively.  They are identical to photons & nucleons, except that
    gluons are labelled as valence partons and, for the nucleon, valence
    quark distributions are set to zero. They are used internally by the
    JIMMY generator for multiple interactions,  and are not intended for
    general use.

  * An error in setting the scale EMCMF (now called EMSCA) for QCD
    decays of colour neutral particles, preventing parton showers, has
    been corrected.

  * Minor bugs have been corrected in:  phi decays to neutral kaons; the
    weights for photo-production processes; the value of EVWGT in di-jet
    production by point-like photons.

  * The transverse momentum cutoff  for final-state photon emission from
    quarks, VPCUT, now defaults to 0.4 GeV.  Previous versions defaulted
    to SQRT(S), switching off such emission.

  * The default top quark mass has been increased to 170 GeV/c^2

------------------------------------------------------------------------

              ****** 3. FEATURES NOT YET INCLUDED ******

   Note that the following features are NOT yet included in the program:
   polarization  of  produced  heavy  quarks  and leptons;  treatment of
   coherence  in the  small-x region  of incoming  jets  (see S. Catani,
   F. Fiorani  and  G. Marchesini,  Nucl.Phys. B336(1990)18);   multiple
   parton  interactions  and  parton  shadowing;  diffractive processes;
   W/Z bosons within parton showers.

------------------------------------------------------------------------

                  ****** 4. PROGRAM STRUCTURE ******

   The main program HWIGPR has the following form:

      PROGRAM HWIGPR
C---COMMON BLOCKS ARE INCLUDED AS FILE HERWIG59.INC
      INCLUDE 'HERWIG59.INC'
      INTEGER N
C---MAX NUMBER OF EVENTS THIS RUN
      MAXEV=100
C---BEAM PARTICLES
      PART1='PBAR'
      PART2='P'
C---BEAM MOMENTA
      PBEAM1=900.
      PBEAM2=900.
C---PROCESS
      IPROC=1500
C---INITIALISE OTHER COMMON BLOCKS
      CALL HWIGIN
C---USER CAN RESET PARAMETERS AT
C   THIS POINT, OTHERWISE DEFAULT
C   VALUES IN HWIGIN WILL BE USED.
      PTMIN=100.
C---COMPUTE PARAMETER-DEPENDENT CONSTANTS
      CALL HWUINC
C---CALL HWUSTA TO MAKE ANY PARTICLE STABLE
      CALL HWUSTA('PI0     ')
C---USER'S INITIAL CALCULATIONS
      CALL HWABEG
C---INITIALISE ELEMENTARY PROCESS
      CALL HWEINI
C---LOOP OVER EVENTS
      DO 100 N=1,MAXEV
C---INITIALISE EVENT
      CALL HWUINE
C---GENERATE HARD SUBPROCESS
      CALL HWEPRO
C---GENERATE PARTON CASCADES
      CALL HWBGEN
C---DO HEAVY QUARK DECAYS
      CALL HWDHQK
C---DO CLUSTER FORMATION
      CALL HWCFOR
C---DO CLUSTER DECAYS
      CALL HWCDEC
C---DO UNSTABLE PARTICLE DECAYS
      CALL HWDHAD
C---DO HEAVY FLAVOUR HADRON DECAYS
      CALL HWDHVY
C---ADD SOFT UNDERLYING EVENT IF NEEDED
      CALL HWMEVT
C---FINISH EVENT
      CALL HWUFNE
C---USER'S EVENT ANALYSIS
      CALL HWANAL
  100 CONTINUE
C---TERMINATE ELEMENTARY PROCESS
      CALL HWEFIN
C---USER'S TERMINAL CALCULATIONS
      CALL HWAEND
      STOP
      END

   Various phases of the  simulation can be  suppressed by  deleting the
   corresponding  subroutine  calls,  or  different  subroutines  may be
   substituted.  For example,  in studies at the parton level everything
   from CALL HWDHQK to CALL HWMEVT can be omitted.

   The following is a full list of subroutines and functions,  which are
   classified  according to their initial letters, except when standard-
   ization agreements take precedence.

       +--------+---------------------------------------------+
       |  Name  | Description                                 |
       +--------+---------------------------------------------+
       |          Main program and initialization             |
       +--------+---------------------------------------------+
       | HWIGPR | Main program                                |
       | HWIGIN | Default initializations                     |
       +--------+---------------------------------------------+
       |          Reading/writing/altering decay modes        |
       +--------+---------------------------------------------+
       | HWIODK | Inputs/outputs formatted decay tables       |
       | HWMODK | Modifies or adds an individual decay mode   |
       +--------+---------------------------------------------+
       |          User-provided analysis routines             |
       +--------+---------------------------------------------+
       | HWABEG | Initializes user's analysis                 |
       | HWAEND | Terminates user's analysis                  |
       | HWANAL | Performs user's analysis on event           |
       +--------+---------------------------------------------+
       |          Parton branching with interfering gluons    |
       +--------+---------------------------------------------+
       | HWBAZF | Computes azimuthal correlation functions    |
       | HWBCON | Makes colour connections between jets       |
       | HWBDED | Correction to the `dead zone' in e+e-       |
       | HWBDIS | Correction to the `dead zone' in DIS        |
       | HWBFIN | Transfers external lines of jet to /HEPEVT/ |
       | HWBGEN | Finds unevolved partons and generates jets  |
       | HWBJCO | Combines jets with correct kinematics       |
       | HWBMAS | Computes masses and trans. momenta in jet   |
       | HWBRAN | Generates a timelike parton branching       |
       | HWBSPA | Computes momenta in spacelike jet           |
       | HWBSPN | Computes spin density/decay matrices        |
       | HWBSU1 | First  term in quark Sudakov form factor    |
       | HWBSU2 | Second term in quark Sudakov form factor    |
       | HWBSUD | Computes (or reads) Sudakov form factors    |
       | HWBSUG | Integrand in gluon Sudakov form factor      |
       | HWBSUL | Logarithmic part of Sudakov form factor     |
       | HWBTIM | Computes momenta in timelike jet            |
       | HWBVMC | Virtual mass cutoff for parton type ID      |
       +--------+---------------------------------------------+
       |          Cluster hadronization model                 |
       +--------+---------------------------------------------+
       | HWCCUT | Cuts a massive cluster in two               |
       | HWCDEC | Decays clusters into primary hadrons        |
       | HWCFLA | Sets up flavours for HWCHAD                 |
       | HWCFOR | Forms clusters                              |
       | HWCGSP | Splits gluons                               |
       | HWCHAD | Decays a cluster into one or two hadrons    |
       +--------+---------------------------------------------+
       |          Particle and heavy quark decays             |
       +--------+---------------------------------------------+
       | HWDBOS | Finds and decays W and Z bosons             |
       | HWDBOZ | Chooses decay mode of W and Z bosons        |
       | HWDCLE | Interface to CLEO package for B decays      |
       | HWDCHK | Checks given decay mode is self-consistent  |
       | HWDFOR | Generates a four-body decay                 |
       | HWDFIV | Generates a five-body decay                 |
       | HWDEUR | Interface to EURODEC package for B decays   |
       | HWDHAD | Generates decays of unstable hadrons        |
       | HWDHGC | Higgs -> gamma gamma decay                  |
       | HWDHGF | Higgs -> W+ W- decay                        |
       | HWDHIG | Finds and decays Higgs bosons               |
       | HWDHQK | Finds and decays heavy quarks               |
       | HWDHVY | Finds and decays heavy flavour hadrons      |
       | HWDIDP | Chooses a parton for HWDHVY                 |
       | HWDPWT | Phase space decay weight                    |
       | HWDTHR | Generates a three-body decay                |
       | HWDTOP | Decides whether to decay top quark          |
       | HWDTWO | Generates a two-body decay                  |
       | HWDWWT | Weak (V-A)  decay weight                    |
       | HWDXLM | Tests if decay vertex lies in given volume  |
       +--------+---------------------------------------------+
       |          Elementary subprocess generation            |
       +--------+---------------------------------------------+
       | HWEFIN | Final calculations on elementary subprocess |
       | HWEGAM | Generates Weizsacker-Williams photon        |
       | HWEINI | Initializes elementary subprocess           |
       | HWEISR | Generates a photon fron initial e or mu     |
       | HWEONE | Sets up a 2->1 hard subprocess              |
       | HWEPRO | Generates elementary subprocess             |
       | HWETWO | Sets up a 2->2 hard subprocess              |
       +--------+---------------------------------------------+
       |          Individual hard subprocesses                |
       +--------+---------------------------------------------+
       | HWHBGF | Hard subprocess: boson-gluon fusion (BGF)   |
       | HWHBKI | Computes kinematics for BGF                 |
       | HWHBRN | Returns a phase-space point for BGF         |
       | HWHBSG | Computes cross section for BGF              |
       | HWHDIS | Hard subprocess: deep inelastic lepton quark|
       | HWHDYP | Hard subprocess: Drell-Yan Z0/photon prodn  |
       | HWHEGG | Hard subprocess: two-photon processes in ee |
       | HWHEGW | Hard subprocess: photon-W processes in e+e- |
       | HWHEGX | Calculates cross section for HWHEGW         |
       | HWHEPA | Hard subprocess: e+e- -> f fbar             |
       | HWHEPG | Hard subprocess: e+e- -> q qbar gluon       |
       | HWHEW0 | e+e- -> W W / Z Z subroutine                |
       | HWHEW1 | e+e- -> W W / Z Z subroutine                |
       | HWHEW2 | e+e- -> W W / Z Z subroutine                |
       | HWHEW3 | e+e- -> W W subroutine                      |
       | HWHEW4 | e+e- -> W W / Z Z subroutine                |
       | HWHEW5 | e+e- -> Z Z subroutine                      |
       | HWHEWW | Hard subprocess: e+e- -> W W / Z Z          |
       | HWHHVY | Hard subprocess: heavy quark production     |
       | HWHIG1 | Matrix elements for Higgs + jet production  |
       | HWHIGA | Amplitudes squared for Higgs + jet          |
       | HWHIGB | Loop integrals for Higgs + jet              |
       | HWHIGJ | QCD Higgs + jet production                  |
       | HWHIGM | Choose Higgs mass for production routines   |
       | HWHIGS | Hard subprocess: gg/qqbar -> Higgs          |
       | HWHIGT | Computes gg -> Higgs cross section          |
       | HWHIGW | Hard subprocess: WW / ZZ -> Higgs           |
       | HWHIGY | Computes ee -> Z -> ZH cross section        |
       | HWHIGZ | Hard subprocess: ee -> Z -> ZH              |
       | HWHPH2 | Hard subprocess: direct photon pairs        |
       | HWHPHO | Hard subprocess: direct photon production   |
       | HWHPPB | Box contribution to gg->photon photon       |
       | HWHPPE | Pointlike photon-parton (fixed flavour)     |
       | HWHPPH | Pointlike photon-parton (fixed pair flavour)|
       | HWHPPM | Pointlike photon-parton direct light meson  |
       | HWHPPT | Pointlike photon-parton (all flavours)      |
       | HWHQPS | Pointlike photon-quark (Compton) scattering |
       | HWHQCD | Hard subprocess: QCD 2->2                   |
       | HWHQCP | Identifies QCD 2->2 hard subprocess         |
       | HWHREM | Treats hard scattering remnants             |
       | HWHSCT | Process extra hard scatterings              |
       | HWHSNG | Colour singlet parton scattering            |
       | HWHSNM | Colour singlet parton scattering ME         |
       | HWHV1J | Hard subprocess W/Z + jet production        |
       | HWHWEX | Top production by W exchange                |
       | HWHWPR | Hard subprocess: W production               |
       +--------+---------------------------------------------+
       |          Soft minimum-bias or underlying event       |
       +--------+---------------------------------------------+
       | HWMEVT | Generates min bias or soft underlying event |
       | HWMLPS | Generates longitudinal phase space          |
       | HWMNBI | Computes negative binomial probability      |
       | HWMULT | Chooses min bias charged multiplicity       |
       | HWMWGT | Calculates weight for minimum bias events   |
       +--------+---------------------------------------------+
       |          Random number generators                    |
       +--------+---------------------------------------------+
       | HWRAZM | Randomly rotated azimuth                    |
       | HWREXP | Random number: exponential distribution     |
       | HWREXQ | Random number: exp. dist. with cutoff       |
       | HWREXT | Random number: exponential transverse mass  |
       | HWRGAU | Random number: Gaussian                     |
       | HWRGEN | Random number generator (l'Ecuyer method)   |
       | HWRINT | Random integer                              |
       | HWRLOG | Random logical                              |
       | HWRPIP | Random primary interaction point            |
       | HWRPOW | Random number: power distribution           |
       | HWRUNG | Random number: uniform + Gaussian tails     |
       | HWRUNI | Random number: uniform                      |
       +--------+---------------------------------------------+
       |          Spacelike branching of incoming partons     |
       +--------+---------------------------------------------+
       | HWSBRN | Generates spacelike parton branching        |
       | HWSDGG | Drees-Grassie photon str. function (gluon)  |
       | HWSDGQ | Drees-Grassie photon str. function (quarks) |
       | HWSFBR | Chooses a spacelike branching               |
       | HWSFUN | Hadron structure functions                  |
       | HWSGAM | Gamma function (for structure functions)    |
       | HWSGEN | Generates x values for spacelike partons    |
       | HWSGQQ | Inserts g->q qbar part of gluon form factor |
       | HWSSPC | Replaces spacelike partons by spectators    |
       | HWSSUD | Sudakov form factor/structure function      |
       | HWSTAB | Interpolates in function table (for HWSSUD) |
       | HWSVAL | Checks for valence parton                   |
       +--------+---------------------------------------------+
       |          Miscellaneous utilities                     |
       +--------+---------------------------------------------+
       | HWUAEM | Running electromagnetic coupling constant   |
       | HWUAER | Real part of photon self-energy             |
       | HWUALF | Two-loop QCD running coupling constant      |
       | HWUANT | Finds a particle's antiparticle             |
       | HWUBPR | Prints branching data for last parton shower|
       | HWUBST | Boost event record to/from hadron-hadron cmf|
       | HWUCFF | Coefficients for e+e- and DIS cross sections|
       | HWUCI2 | Logarithmic integral Ci_2                   |
       | HWUDAT | Block data: particle properties             |
       | HWUDKL | Generates decay vertex of unstable particle |
       | HWUDKS | Converts decay modes into internal format   |
       | HWUDPR | Prints particle properties and decay modes  |
       | HWUECM | Centre-of-mass energy                       |
       | HWUEDT | Insert or delete entries in the event record|
       | HWUEEC | Computes coefficients for e+e- cross section|
       | HWUEPR | Prints event data                           |
       | HWUEMV | Moves entries within the event record       |
       | HWUFNE | Finishes an event                           |
       | HWUGAU | Adaptive Gaussian integration               |
       | HWUIDT | Translates particle identity codes          |
       | HWUINC | Initial parameter-dependent calculations    |
       | HWUINE | Initializes an event                        |
       | HWULB4 | Boost: rest frame -> lab, no masses assumed |
       | HWULDO | Lorentz 4-vector dot product                |
       | HWULF4 | Boost: lab frame -> rest, no masses assumed |
       | HWULI2 | Logarithmic integral Li_2 (Spence function) |
       | HWULOB | Lorentz transformation: rest frame -> lab   |
       | HWULOF | Lorentz transformation: lab -> rest frame   |
       | HWULOR | Multiplies by Lorentz matrix                |
       | HWUMAS | Puts mass in 5th component of vector        |
       | HWUPCM | Centre-of-mass momentum                     |
       | HWURAP | Rapidity                                    |
       | HWURES | Computes/prints resonance data              |
       | HWUROB | Rotation by inverse of matrix R             |
       | HWUROF | Rotation by matrix R                        |
       | HWUROT | Computes rotation R from vector to z-axis   |
       | HWUSOR | Sorts an array in ascending order           |
       | HWUSQR | Square root with sign retention             |
       | HWUSTA | Makes a particle type stable                |
       | HWUTAB | Interpolates in a table                     |
       | HWUTIM | Checks time remaining (N.B. VAX Fortran)    |
       +--------+---------------------------------------------+
       |          Vector manipulation                         |
       +--------+---------------------------------------------+
       | HWVDIF | Vector difference                           |
       | HWVDOT | Vector dot product                          |
       | HWVEQU | Vector equality                             |
       | HWVSCA | Vector times scalar                         |
       | HWVSUM | Vector sum                                  |
       | HWVZRO | Vector zero                                 |
       +--------+---------------------------------------------+
       |          Warning messages and error handling         |
       +--------+---------------------------------------------+
       | HWWARN | Issues warnings and deals with errors       |
       +--------+---------------------------------------------+

   N.B. Dummy versions of the external routines

   PDFSET  STRUCTM
   EUDINI  FRAGMT  IEUPDG  IPDGEU
   DECADD  QQINIT  QQLMAT
   HVCBVI  HVHBVI

   should be deleted if the structure function library,  EURODEC B decay
   package,  CLEO B decay package,  or HERBVI (respectively) is linked.
------------------------------------------------------------------------

                 ****** 5. BEAMS AND PROCESSES ******

   As  indicated  above,  a  number of variables must be set in the main
   program to specify what is to be simulated:

        +----------+----------------------------------+-----------+
        |   Name   |     Description                  |  Default  |
        +----------+----------------------------------+-----------+
        | PART1    | Type of particle in beam 1       | 'PBAR    '|
        | PART2    | Type of particle in beam 2       | 'P       '|
        | PBEAM1   | Momentum of beam 1               | 900.      |
        | PBEAM2   | Momentum of beam 2               | 900.      |
        | IPROC    | Type of process to generate      | 1500      |
        | MAXEV    | Number of events to generate     | 100       |
        +----------+----------------------------------+-----------+

    The beam particle types PART1,PART2 supported at present are:

              +---------------------------------------------+
              | 'E+      ','E-      ','MU+     ','MU-     ' |
              | 'NUE     ','NUEB    ','NUMU    ','NMUB    ' |
              | 'NTAU    ','NTAB    ','GAMA    '            |
              | 'P       ','PBAR    ','N       ','NBAR    ' |
              | 'PI+     ','PI-     '                       |
              +---------------------------------------------+

    In  addition,  beams 'K+      '  and  'K-      ' are  supported  for
    minimum bias non-diffractive soft hadronic events (IPROC=8000) only.

    The currently available processes IPROC are tabulated below.

    +---------+--------------------------------------------------------+
    |  IPROC  |                     Process                            |
    +---------+--------------------------------------------------------+
    | 100     | e+ e-  -> q qbar (gluon) (all flavours)                |
    | 100+IQ  | e+ e-  -> q qbar (gluon) (IQ=1--6 for q=d,u,s,c,b,t)   |
    | 107     | e+ e-  -> gluon gluon (gluon) fictitious process       |
    | 110     | e+ e-  -> q qbar gluon (all flavours)                  |
    | 110+IQ  | e+ e-  -> q qbar gluon (IQ as above)                   |
    | 120     | e+ e-  -> q qbar (all flavours)| without correction to |
    | 120+IQ  | e+ e-  -> q qbar (IQ as above) | hard gluon branching  |
    | 127     | e+ e-  -> gluon gluon          |                       |
    | 150+IL  | e+ e-  -> l lbar (IL=2,3 for l=mu,tau)                 |
    +---------+--------------------------------------------------------+
    | 200     | e+ e-  -> W+ W- (see sect. 18 on control of W/Z decays)|
    | 250     | e+ e-  -> Z0 Z0 (see sect. 18 on control of W/Z decays)|
    +---------+--------------------------------------------------------+
    | 300     | e+ e-  -> Z H -> Z q qbar (all flavours)               |
    | 300+IQ  | e+ e-  -> Z H -> Z q qbar (IQ as above)                |
    | 306+IL  | e+ e-  -> Z H -> Z l lbar (IL=1,2,3 for l=e,mu,tau)    |
    | 310,11  | e+ e-  -> Z H -> Z W W, Z Z Z                          |
    | 312     | e+ e-  -> Z H -> Z gamma gamma                         |
    | 399     | e+ e-  -> Z H -> Z anything                            |
    +---------+--------------------------------------------------------+
    | 400+ID  | e+ e-  -> nu nu H + e e H (ID as in IPROC=300+ID)      |
    +---------+--------------------------------------------------------+
    | 500+ID  | e+ e-  -> gamma gamma -> qqbar/llbar/WW (ID=0-10 as in |
    |         |                                          IPROC=300+ID) |
    | 550+ID  | e+ e-  -> gamma W -> qq'bar/ll'bar      (ID=0-9)       |
    +---------+--------------------------------------------------------+
    | 1300    | q qbar -> Z0/gamma -> q qbar (all flavours)            |
    | 1300+IQ | q qbar -> Z0/gamma -> q qbar (IQ as above)             |
    | 1350    | q qbar -> Z0/gamma -> l lbar (all lepton species)      |
    | 1350+IL | q qbar -> Z0/gamma -> l lbar (IL=1-6 for e,enu,mu,etc) |
    | 1399    | q qbar -> Z0/gamma -> anything                         |
    +---------+--------------------------------------------------------+
    | 1400    | q qbar -> W+/- -> q' qbar'' (all flavours)             |
    | 1400+IQ | q qbar -> W+/- -> q' qbar'' (q' or q'' as above)       |
    | 1450    | q qbar -> W+/- -> l nul (all lepton species)           |
    | 1450+IL | q qbar -> W+/- -> l nul (IL=1-3 as above)              |
    | 1499    | q qbar -> W+/- -> anything                             |
    +---------+--------------------------------------------------------+
    | 1500    | QCD  2 -> 2 hard parton scattering                     |
    |         | After generation, IHPRO is subprocess (see list)       |
    +---------+--------------------------------------------------------+
    | 1600+ID | q qbar/g g -> Higgs (ID as in IPROC=300+ID)            |
    +---------+--------------------------------------------------------+
    | 1700+IQ | QCD heavy quark production (IQ as above)               |
    |         | After generation, IHPRO is subprocess (see list)       |
    +---------+--------------------------------------------------------+
    | 1800    | QCD direct photon + jet production                     |
    |         | After generation, IHPRO is subprocess (see list)       |
    +---------+--------------------------------------------------------+
    | 1900+ID | q qbar -> q' qbar' H (ID as in IPROC=300+ID)           |
    +---------+--------------------------------------------------------+
    | 2000    |    t production via W exchange (sum of 2001-2008)      |
    | 2001,2  |    ubar bbar -> dbar tbar,   d bbar -> u tbar          |
    | 2003,4  |    dbar bbar -> ubar tbar,   u b    -> d t             |
    | 2005,6  |    cbar bbar -> sbar tbar,   s bbar -> c tbar          |
    | 2007,8  |    sbar b    -> cbar t   ,   c b    -> s t             |
    +---------+--------------------------------------------------------+
    | 2100    | Vector boson + jet production.                         |
    | 2110,20 | Compton only (g q -> V q), annih. only (q qbar -> V g) |
    +---------+--------------------------------------------------------+
    | 2200    | QCD direct photon pair production (see list for IHPRO) |
    +---------+--------------------------------------------------------+
    | 2300    | QCD Higgs plus jet production (see list for IHPRO)     |
    +---------+--------------------------------------------------------+
    | 2400    | Mueller-Tang colour singlet exchange                   |
    | 2450    | Quark scattering via photon exchange                   |
    +---------+--------------------------------------------------------+
    | 5000    | Pointlike photon-hadron jet production (all flavours)  |
    | 5100+IQ | Pointlike photon heavy flavour IQ pair production      |
    | 5200+IQ | Pointlike photon heavy flavour IQ single excitation    |
    |         | After generation, IHPRO is subprocess (see list)       |
    | 5300    | Quark photon Compton scattering                        |
    | 5500    | Pointlike photon production of light (u,d,s) L=0 mesons|
    | 5510,20 | S=0 mesons only, S=1 mesons only (see list for IHPRO)  |
    +---------+--------------------------------------------------------+
    | 7000 -  | Baryon-number violating and other multi-W processes    |
    | 7999    | generated by HERBVI package                            |
    +---------+--------------------------------------------------------+
    | 8000    | Minimum bias non-diffractive soft hadron-hadron event  |
    +---------+--------------------------------------------------------+
    | 9000    | Deep inelastic lepton scattering (all neutral current) |
    | 9000+IQ | Deep inelastic lepton scattering (NC on flavour IQ)    |
    | 9010    | Deep inelastic lepton scattering (all charged current) |
    | 9010+IQ | Deep inelastic lepton scattering (CC on flavour IQ)    |
    +---------+--------------------------------------------------------+
    | 9100+IQ | Boson-gluon fusion in NC DIS, IQ=0-6 as above          |
    | 9107    | J/Psi + gluon production by boson-gluon fusion         |
    | 9110    | QCD Compton process in NC DIS, all flavours            |
    | 9110+IP | QCD Compton process in NC DIS, IP=1-12, d-t, dbar-tbar |
    | 9130    | All O(alpha-s) NC processes: 9100+9110                 |
    | 9140+IP | CC proc, IP:1 = s cbar,2 = b cbar,3 = s tbar,4 = b tbar|
    +---------+--------------------------------------------------------+
    | 92**    | Withdrawn: use 91** instead                            |
    +---------+--------------------------------------------------------+
    | 9500+ID | W W fusion -> Higgs in e p (ID as in IPROC=300+ID)     |
    +---------+--------------------------------------------------------+
    |10000+IP | as IPROC=IP but with soft underlying event (hadron     |
    |         | remnant fragmentation in lepton-hadron) suppressed     |
    +---------+--------------------------------------------------------+

     The  extent  to  which  quark mass effects are included in the hard
     process cross section is different in different processes.  In many
     processes,  they are always treated as massless:  IPROC=1300, 1800,
     1900,  2100, 2300, 2400, 5300, 9000.  In two processes they are all
     treated  as  massive  except  the  top quark, for which the mass is
     correctly  incorporated:  1400, 2000.  In the case of massless pair
     production,  only quark flavours that are kinematically allowed are
     produced.   In all cases the event kinematics incorporate the quark
     mass, even when it is not used to calculate the cross section.

     In two processes, quarks are always treated as massive: 500, 9100.

     Finally, in several processes, the behaviour is different depending
     on whether a specific quark flavour is requested, in which case its
     mass  is  included, or not, in which case all quarks are treated as
     massless.   These  are:  IPROC=100,  110,  120, QCD 2->2 scattering
     (1500  vs 1700+IQ), jets in direct photoproduction (5000 vs 5100+IQ
     and 5200+IQ).

     These  differences can cause inconsistencies between different ways
     of  generating the same process.  The most noticeable example is in
     direct  photoproduction, where one can use process 9130, which uses
     the  exact  2->3  matrix element e+g --> e+q+qbar, or process 5000,
     which  uses  the Weizsacker-Williams spectrum for e --> e+gamma and
     the  2->2  matrix element for gamma+g --> q+qbar.  For typical HERA
     kinematics,  the  W-W approximation is valid to a few per cent, but
     the  difference between the two processes is much larger, about 20%
     for  PTMIN=2 GeV.   This is entirely due to the difference in quark
     mass  treatments,  as can be checked by comparing process 9130 with
     processes 5100+IQ and 5200+IQ summed over IQ
------------------------------------------------------------------------

                   ****** 6. INPUT PARAMETERS ******

   The quantities that may be regarded as adjustable parameters are

         +----------+----------------------------------+-------+
         |   Name   |     Description                  |Default|
         +----------+----------------------------------+-------+
         | QCDLAM   | QCD Lambda (see below)           | 0.18  |
         +----------+----------------------------------+-------+
         | RMASS(1) | Down    quark mass               | 0.32  |
         | RMASS(2) | Up      quark mass               | 0.32  |
         | RMASS(3) | Strange quark mass               | 0.50  |
         | RMASS(4) | Charmed quark mass               | 1.55  |
         | RMASS(5) | Bottom  quark mass               | 4.95  |
         | RMASS(6) | Top     quark mass               | 170.  |
         +----------+----------------------------------+-------+
         | RMASS(13)| Gluon effective mass             | 0.75  |
         +----------+----------------------------------+-------+
         | VQCUT    | Quark virtuality cutoff (added to| 0.48  |
         |          | quark masses in parton showers)  |       |
         | VGCUT    | Gluon virtuality cutoff (added to| 0.10  |
         |          | effective mass in parton showers)|       |
         | VPCUT    | Photon virtuality cutoff         | 0.40  |
         +----------+----------------------------------+-------+
         | CLMAX    | Maximum cluster mass parameter   | 3.35  |
         | CLPOW    | Power in maximum cluster mass    | 2.00  |
         | PSPLT    | Split cluster spectrum parameter | 1.00  |
         +----------+----------------------------------+-------+
         | QDIQK    | Maximum scale for gluon->diquarks| 0.00  |
         | PDIQK    | Gluon->diquarks rate parameter   | 5.00  |
         +----------+----------------------------------+-------+
         | QSPAC    | Cutoff for spacelike evolution   | 2.50  |
         | PTRMS    | Intrinsic pt in incoming hadrons | 0.00  |
         +----------+----------------------------------+-------+

   Notes on parameters:

  *  QCDLAM can be identified  at high momentum fractions  (x or z) with
     the fundamental QCD scale Lambda-MSbar (5 flavours).  However, this
     relation does not necessarily hold in other regions of phase space,
     since higher order corrections are not treated  precisely enough to
     remove renormalization scheme ambiguities. See S. Catani, G. March-
     esini and B.R.Webber, Nucl. Phys. B349 (1991) 635.

  *  RMASS(1,2,3,13)  are effective light quark and gluon masses used in
     the  hadronization phase  of the program.  They can be  set to zero
     provided the parton shower cutoffs VQCUT and VGCUT are large enough
     to prevent divergences (see below).

  *  For cluster hadronization, it must be possible to split gluons into
     q-qbar, i.e. RMASS(13)  must be at  least twice the  lightest quark
     mass.  Similarly it may be  impossible for heavy flavoured clusters
     to decay if RMASS(4,5) are too low.

  *  VQCUT and VGCUT are needed if the  quark and gluon effective masses
     become small. The condition to avoid divergences in  parton showers
     is
         1/Q(i) + 1/Q(j) < 1/QCDL3  for either i or j or both gluons,
     where  Q(i)=RMASS(i)+VQCUT for quarks,  RMASS(13)+VGCUT for gluons,
     and QCDL3 is the equivalent 3-flavour Lambda computed from  QCDLAM.
     In the notation of  the above reference by  S. Catani et al., QCDL3
     is the 3-flavour equivalent of QCDL5 where
           QCDL5 = QCDLAM*exp(K/(4*pi*beta))/sqrt(2)=1.109*QCDLAM

  *  VPCUT is the analogous quantity for photon emission. It defaults to
     SQRT(S)  corresponding to no emission.  Results after  experimental
     cuts are insensitive to its exact value in the range 0.1 to 1.0 GeV

  *  CLMAX and CLPOW  determine the  maximum allowed  mass of  a cluster
     made from quarks i and j as follows
           Mass**CLPOW < CLMAX**CLPOW + (RMASS(i)+RMASS(j))**CLPOW
     Since the cluster mass spectrum falls rapidly at high mass, results
     become  insensitive to  CLMAX and CLPOW  at large  values of CLMAX.
     Smaller values OF CLPOW will increase the yield of heavier clusters
     (and hence of baryons)  for heavy quarks,  without affecting  light
     quarks  much.  For example,  the default  value gives  no b-baryons
     whereas CLPOW=1.0 makes b-baryons/b-hadrons about 1/4.

  *  PSPLT  determines the  mass  distribution in the  cluster splitting
     CL1 -> CL2 + CL3  when CL1 is above  the maximum allowed mass.  The
     masses of CL2 and CL3 are generated uniformly in Mass**PSPLT. Since
     the number of split clusters is small, dependence on PSPLT is weak.

  *  QDIQK greater than twice the  lightest diquark mass  enables gluons
     to split  non-perturbatively into  diquarks as well as quarks.  The
     probability  of this is  PDIQK*dQ/Q  for scales Q  below QDIQK. The
     diquark masses are taken to be the sum of constituent quark masses.
     Thus the default value QDIQK=0 suppresses gluon->diquark splitting.

  *  QSPAC is the scale below which the structure functions of incoming
     hadrons are frozen and non-valence constituent partons are forced
     to evolve to valence partons, if ISPAC=0.  For ISPAC=2, structure
     functions are frozen at scale QSPAC, but evolution continues down
     to the infrared cutoff.

  *  PTRMS is the width of the (Gaussian) intrinsic  transverse momentum
     distribution of valence partons in incoming hadrons at scale QSPAC.
     (N.B. Neither  QSPAC nor PTRMS  affect  lepton-lepton  collisions.)

   In practice, the parameters  that have been  found most  effective in
   fitting data are QCDLAM,  the gluon effective mass RMASS(13), and the
   cluster mass parameter CLMAX.

   The default parameter values  have been found to give  good agreement
   with event shape  distributions at LEP (OPAL preprint CERN-EP/90-48).

   A number of  further parameters are needed to control the program and
   to turn various options on or off:

         +----------+----------------------------------+-------+
         |   Name   |     Description                  |Default|
         +----------+----------------------------------+-------+
         | IPRINT   | Printout option                  | 1     |
         | MAXPR    | Number of events to print out    | 1     |
         | PRVTX    | Include vertex info in print out | .TRUE.|
         | MAXER    | Max number of errors             | 10    |
         | LWEVT    | Unit for writing output events   | 0     |
         | LRSUD    | Unit for reading Sudakov table   | 0     |
         | LWSUD    | Unit for writing Sudakov table   | 77    |
         | SUDORD   | Alpha_s order in Sudakov table   | 1     |
         +----------+----------------------------------+-------+
         | NRN(1)   | Random number seed 1             | 17673 |
         | NRN(2)   | Random number seed 2             | 63565 |
         | WGTMAX   | Max weight (0 to search for it)  | 0.    |
         | NOWGT    | Generate unweighted events       | .TRUE.|
         +----------+----------------------------------+-------+
         | AZSOFT   | Soft gluon azimuthal correlations| .TRUE.|
         | AZSPIN   | Gluon spin azimuthal correlations| .TRUE.|
         +----------+----------------------------------+-------+
         | NCOLO    | Number of colours                | 3     |
         | NFLAV    | Number of (producible) flavours  | 6     |
         +----------+----------------------------------+-------+
         | MODPDF(I)| PDFLIB structure function set and| -1    |
         | AUTPDF(I)| author group for beam I(=1,2)    | 'MRS' |
         |          | (if MODPDF()<0 do not use PDFLIB)|       |
         | NSTRU    | Input structure function set     | 5     |
         |          | (1,2=Duke-Owens1,2   3,4=EHLQ1,2 |       |
         |          |    5=Owens1.1)                   |       |
         +----------+----------------------------------+-------+
         | ETAMIX   | eta/eta' mixing angle in degrees | -20   |
         |          | F0Mix..
         +----------+----------------------------------+-------+
         | B1LIM    | B cluster -> 1 hadron parameter  |  0.0  |
         +----------+----------------------------------+-------+
         | CLDIR    | Decay of perturbative clusters,  | 1     |
         |          | 0=>isotropic, 1=>along quark dirn|       |
         | CLSMR    | Width of Gaussian angle smearing | 0.0   |
         +----------+----------------------------------+-------+
         | CLRECO   | Include colour rearrangement     |.FALSE |
         | PRECO    | Probability for rearrangement    | 1./9. |
         | EXAG     | Lifetime scaling for weak bosons | 1.    |
         +----------+----------------------------------+-------+
         | PIPSMR   | Smear the primary vertex         | .TRUE.|
         | MAXDKL   | Veto decays outside given volume |.FALSE.|
         +----------+----------------------------------+-------+
         | HARDME   | Use hard and soft matrix-element | .TRUE.|
         | SOFTME   | corrections to e+e- and DIS      | .TRUE.|
         +----------+----------------------------------+-------+
         | BDECAY   | Controls which B Decay package is| 'HERW'|
         |          | used. The allowed values are:    |       |
         |          | 'HERW'; 'EURO'; or 'CLEO'.       |       |
         | MIXING   | Include neutral B meson mixing   | .TRUE.|
         | XMIX(2)  | Mass difference     I=1 B^0_s    | 10.0  |
         |          |  average width        2 B^0_d    | 0.70  |
         | YMIX(2)  | Width difference    I=1 B^0_s    | 0.20  |
         |          |  average width        2 B^0_d    | 0.00  |
         +----------+----------------------------------+-------+
         | EPOLN(3) | Electron and positron beam       | 0.0   |
         |          | polarizations in DIS and e+e-    | 0.0   |
         |          | annihilation. First two cmpts are| 0.0   |
         | PPOLN(3) | transverse and only used in e+e-,| 0.0   |
         |          | 3rd cmpt is longitudinal, and is | 0.0   |
         |          | +/-1 for fully rh/lh polarized   | 0.0   |
         +----------+----------------------------------+-------+
         | BGSHAT   | Scale=shat for boson-gluon fusion|.FALSE.|
         +----------+----------------------------------+-------+
         | BREIT    | Use Breit frame for DIS kinematix| .TRUE.|
         +----------+----------------------------------+-------+
         | USECMF   | Use hadron-hadron cmf            | .TRUE.|
         +----------+----------------------------------+-------+
         | NOSPAC   | Switch off space-like showers    |.FALSE.|
         +----------+----------------------------------+-------+
         | ISPAC    | Changes meaning of QSPAC,        | 0     |
         |          | see the earlier notes on QSPAC   |       |
         +----------+----------------------------------+-------+
         | TMNISR   | Min vaule shat/S for photon ISR  | 1D-4  |
         | ZMNISR   | Max mom fraction for photon ISR  | 1-1D-6|
         +----------+----------------------------------+-------+
         | PTMIN    | Min pt in hadronic jet production| 10.   |
         | PTMAX    | Max pt in hadronic jet production| 1.E8  |
         | PTPOW    | 1/pt**PTPOW for jet sampling     | 4.    |
         | YJMIN    | Min jet rapidity                 |-8.    |
         | YJMAX    | Max jet rapidity                 | 8.    |
         +----------+----------------------------------+-------+
         | EMMIN    | Min dilepton mass in Drell-Yan   | 10.   |
         | EMMAX    | Max dilepton mass in Drell-Yan   | 1.E8  |
         | EMPOW    | 1/m**EMPOW for Drell-Yan sampling| 4.    |
         +----------+----------------------------------+-------+
         | Q2MIN    | Min Q**2 in deep inelastic       | 0.0   |
         | Q2MAX    | Max Q**2 in deep inelastic       | 1.E10 |
         | Q2POW    | (1/Q**2)**Q2POW for sampling     | 2.5   |
         +----------+----------------------------------+-------+
         | Q2WWMN   | Min Q**2 in Equiv Photon Approx  | 0.0   |
         | Q2WWMX   | Max Q**2 in Equiv Photon Approx  | 4.0   |
         +----------+----------------------------------+-------+
         | YWWMIN   | Min energy of gamma in WW approx | 1.0   |
         | YWWMAX   | Max energy of gamma in WW approx | 0.0   |
         +----------+----------------------------------+-------+
         | PHOMAS   | Damp structure functions for off-| 0.0   |
         |          | shell photons (0 for no damping) |       |
         +----------+----------------------------------+-------+
         | YBMIN    | Min and Max Bjorken-y in DIS and | 0.0   |
         | YBMAX    | Equivalent Photon Approx         | 1.0   |
         +----------+----------------------------------+-------+
         | ZJMAX    | Max Z in J/psi production        | 0.9   |
         +----------+----------------------------------+-------+
         | THMAX    | Max thrust in 3 parton production| 0.9   |
         |          | (equal to 1-Y_cut in JADE scheme)|       |
         +----------+----------------------------------+-------+

     Printout options are:

           IPRINT = 0   Print program title only
                    1   Print selected input parameters
                    2   1 + table of particle codes and properties
                    3   2 + tables of Sudakov form factors

           PRVTX = .T. To include the  production vertex information in
                       the event print out, requires wide screen format.

     See sect. 8 on form factors for details of LRSUD, LWSUD and SUDORD.

     If BGSHAT is false, the scale used for heavy quark production via
     boson-gluon fusion in lepton-hadron collisions will be
             2*shat*that*uhat/(shat**2+that**2+uhat**2)

     If BREIT is true,  the kinematic reconstruction of  deep inelastic
     events  takes place  in the Breit frame  (ie. the frame  where the
     exchanged  boson  is  purely  space-like,  and collinear  with the
     incoming  hadron).    In  fact  the  reconstruction  procedure  is
     invariant  under  longitudinal  boosts,  so any frame in which the
     boson and hadron are collinear would be equivalent, and it is only
     the transverse part of the boost that has an effect.
     The BREIT frame option becomes very inaccurate for very small Q^2.
     It is therefore only used if Q**2 > 1E-4 (the lab and Breit frames
     are anyway equivalent for such small Q**2).
     If BREIT is false, reconstruction takes place in the lab frame.

     If USECMF is true, the entire event record is boost to the hadron-
     hadron cmf before  event processing,  and boosted back afterwards.
     This means that  fixed-target simulation  can be done  in the  lab
     frame, ie with PBEAM2=0.
     For hadronic processes with lepton beams,  this boosting is always
     done, regardless of the value of USECMF.

     The  interface  to  the  PDFLIB  structure  function  package  is
     compatible with PDFLIB versions 3 and 4.  For version 4, AUTPDF()
     should be set to the author group as listed in the PDFLIB manual,
     eg  'MRS',  and MODPDF() to the set number in the new convention.
     For version 3,  AUTPDF() should be set to 'MODE', and MODPDF() to
     the set number in the old convention.

     The `hard' matrix-element correction adds e+e- and DIS events in
     regions of phase-space that cannot be filled by the usual parton
     shower.  The  `soft'  matrix-element  correction moves emissions
     around within the shower  phase-space,   essentially by matching
     the HARDEST emission (which is not necessarily the first) to the
     first-order matrix-element.

     The quantities  from  PTMIN  onwards  control the region of  phase
     space in  which events  are generated and  the importance sampling
     inside those regions.  See section 11 on event weights for further
     details on  these  quantities  and the use of  WGTMAX  and  NOWGT.

     If hadronic processes with lepton beams are requested,  the photon
     emission vertex  includes the  full  transverse-momentum-dependent
     kinematics  (the Equivalent Photon Approximation).   The variables
     Q2WWMN  and  Q2WWMX  set  the  minimum  and  maximum  virtualities
     generated respectively. For normal simulation, Q2WWMN should be 0,
     and Q2WWMX should be the largest Q**2 through which the lepton can
     be scattered  without  being  detected.  The  variables  YBMIN and
     YBMAX control the range of lightcone momentum fraction generated.

     In addition  there are  options to  give  different weights to the
     various  flavours  of quarks and  diquarks,  and to  resonances of
     different spins.  So far,  these options  have not  been used. See
     the comments in the  initialization  routine  HWIGIN  for details.

------------------------------------------------------------------------

                  ****** 7. COMMON BLOCK FILE ******

C          ****COMMON BLOCK FILE FOR HERWIG VERSION 5.9****
C
C ALTERATIONS: See 5.8 for list of previous revisions
C              Layout completely overhauled
C
C The following variables have been removed:
C              FBTM,FTOP,FHVY,VECWT,TENWT,SWT,RESWT
C              MADDR,MODES,MODEF,IDPRO
C The following COMMON BLOCK has been removed
C              /HWUFHV/   - BDECAY moved to /HWPRCH/
C The following COMMON BLOCKs have been added
C              /HWBMCH/   -contains PART1, PART2 from /HWBEAM/
C              /HWPRCH/   -contains AUTPDF from /HWPARM/ & BDECAY         
C              /HWPROP/   -contains many variables from /HWUPDT/
C              /HWDIST/   -contains variables for mixing and vertices
C              /HWQDKS/   -contains heavy flavour decay information
C The following variables have been changed to CHARACTER*8:
C              PART1,PART2,RNAME
C The following parameters have been added:
C              NMXCDK,NMXDKS,NMXMOD,NMXQDK,NMXRES
C The following variables have been added:
C              CSPEED,F0MIX,F1MIX,F2MIX,H1MIX,
C              PHIMIX,IOPREM,PRVTX                 see HWPRAM
C              ANOMSC,ISLENT                       see HWBRCH
C              GAMWT                               see HWEVNT
C              ASFIXD,OMEGA0,TMNISR,WHMIN,YWWMAX,  
C              YWWMIN,ZMXISR,COLISR                see HWHARD
C              IFLAV,RLTIM,RSPIN,VTOCDK,VTORDK     see HWPROP
C              DKLTM,IDK,IDKPRD,LNEXT,LSTRT,
C              NDKYS,NME,NMODES,NPRODS,
C              DKPSET,RSTAB                        see HWUPDT
C              REPWT,SNGWT                         see HWUWTS
C              CLDKWT,CTHRPW,PRECO,NCLDK,CLRECO    see HWUCLU
C              EXAG,GEV2MM,HBAR,PLTCUT,VMIN2,
C              VTXPIP,XMIX,XMRCT,YMIX,YMRCT,
C              IOPDKL,MAXDKL,MIXING,PIPSMR         see HWDIST
C              VTXQDK,IMQDK,LOCQ,NQDK              see HWQDKS
C
C
      IMPLICIT NONE
      DOUBLE PRECISION ZERO,ONE,TWO,THREE,FOUR,HALF
      PARAMETER (ZERO =0.D0, ONE =1.D0, TWO =2.D0,
     &           THREE=3.D0, FOUR=4.D0, HALF=0.5D0)
C
      DOUBLE PRECISION
     & ACCUR,AFCH,ALPFAC,ALPHEM,ANOMSC,ASFIXD,AVWGT,B1LIM,BETAF,BRFRAC,
     & BRHIG,BTCLM,CAFAC,CFFAC,CLDKWT,CLMAX,CLPOW,CLQ,CLSMR,CMMOM,COSS,
     & COSTH,CSPEED,CTHRPW,CTMAX,DECPAR,DECWT,DISF,DKLTM,EBEAM1,EBEAM2,
     & EMLST,EMMAX,EMMIN,EMPOW,EMSCA,ENHANC,ENSOF,EPOLN,ETAMIX,EVWGT,
     & EXAG,F0MIX,F1MIX,F2MIX,GAMH,GAMMAX,GAMW,GAMWT,GAMZ,GAMZP,GCOEF,
     & GEV2NB,GEV2MM,GPOLN,H1MIX,HBAR,HARDST,OMEGA0,PBEAM1,PBEAM2,PDIQK,
     & PGSMX,PGSPL,PHEP,PHIMIX,PHIPAR,PHOMAS,PIFAC,PLTCUT,PPAR,PPOLN,
     & PRECO,PRSOF,PSPLT,PTINT,PTMAX,PTMIN,PTPOW,PTRMS,PXRMS,PWT,Q2MAX,
     & Q2MIN,Q2POW,Q2WWMN,Q2WWMX,QCDL3,QCDL5,QCDLAM,QDIQK,QEV,QFCH,QG,
     & QLIM,QSPAC,QV,QWT,REPWT,RESN,RHOHEP,RHOPAR,RLTIM,RMASS,RMIN,
     & RSPIN,SCABI,SINS,SNGWT,SWEIN,SWTEF,SUD,THMAX,TLOUT,TMTOP,TMNISR,
     & TQWT,VCKM,VFCH,VGCUT,VHEP,VMIN2,VPAR,VPCUT,VQCUT,VTXPIP,VTXQDK,
     & WBIGST,WGTMAX,WGTSUM,WHMIN,WSQSUM,XFACT,XLMIN,XMIX,XMRCT,XX,
     & XXMIN,YBMAX,YBMIN,YJMAX,YJMIN,YMIX,YMRCT,YWWMAX,YWWMIN,ZBINM,
     & ZJMAX,ZMXISR
C
      INTEGER
     & CLDIR,IAPHIG,IBRN,IBSH,ICHRG,ICO,IDCMF,IDHEP,IDHW,IDK,IDKPRD,IDN,
     & IDPAR,IDPDG,IERROR,IFLAV,IFLMAX,IFLMIN,IHPRO,IMQDK,INHAD,INTER,
     & IOPDKL,IOPHIG,IOPREM,IPART1,IPART2,IPRINT,IPRO,IPROC,ISLENT,
     & ISPAC,ISTAT,ISTHEP,ISTPAR,JCOPAR,JDAHEP,JDAPAR,JMOHEP,JMOPAR,
     & JNHAD,LNEXT,LOCN,LOCQ,LRSUD,LSTRT,LWEVT,LWSUD,MAPQ,MAXER,MAXEV,
     & MAXFL,MAXPR,MODBOS,MODMAX,MODPDF,NBTRY,NCLDK,NCOLO,NCTRY,NDKYS,
     & NDTRY,NETRY,NEVHEP,NEVPAR,NFLAV,NGSPL,NHEP,NME,NMODES,NMXCDK,
     & NMXDKS,NMXHEP,NMXJET,NMXMOD,NMXPAR,NMXQDK,NMXRES,NMXSUD,NPAR,
     & NPRODS,NQDK,NQEV,NRES,NRN,NSPAC,NSTRU,NSTRY,NSUD,NUMER,NUMERU,
     & NWGTS,NZBIN,SUDORD
C
      LOGICAL
     & AZSOFT,AZSPIN,BGSHAT,BREIT,CLRECO,COLISR,DKPSET,FROST,FSTEVT,
     & FSTWGT,GENEV,GENSOF,HARDME,HVFCEN,MAXDKL,MIXING,NOSPAC,NOWGT,
     & PRNDEC,PIPSMR,PRVTX,RSTAB,SOFTME,TMPAR,TPOL,USECMF,VTOCDK,VTORDK,
     & ZPRIME
C
      CHARACTER*4
     & BDECAY
      CHARACTER*8
     & PART1,PART2,RNAME
      CHARACTER*20
     & AUTPDF
C
C New standard event common
      PARAMETER (NMXHEP=2000)
      COMMON/HEPEVT/NEVHEP,NHEP,ISTHEP(NMXHEP),IDHEP(NMXHEP),
     & JMOHEP(2,NMXHEP),JDAHEP(2,NMXHEP),PHEP(5,NMXHEP),VHEP(4,NMXHEP)
C
C Beams, process and number of events
      COMMON/HWBEAM/IPART1,IPART2
      COMMON/HWBMCH/PART1,PART2
      COMMON/HWPROC/EBEAM1,EBEAM2,PBEAM1,PBEAM2,IPROC,MAXEV
C
C Basic parameters (and quantities derived from them)
      COMMON/HWPRAM/AFCH(16,2),ALPHEM,B1LIM,BETAF,BTCLM,CAFAC,CFFAC,
     & CLMAX,CLPOW,CLSMR,CSPEED,ENSOF,ETAMIX,F0MIX,F1MIX,F2MIX,GAMH,
     & GAMW,GAMZ,GAMZP,GEV2NB,H1MIX,PDIQK,PGSMX,PGSPL(4),PHIMIX,PIFAC,
     & PRSOF,PSPLT,PTRMS,PXRMS,QCDL3,QCDL5,QCDLAM,QDIQK,QFCH(16),QG,
     & QSPAC,QV,SCABI,SWEIN,TMTOP,VFCH(16,2),VCKM(3,3),VGCUT,VQCUT,
     & VPCUT,ZBINM,IOPREM,IPRINT,ISPAC,LRSUD,LWSUD,MODPDF(2),NBTRY,
     & NCOLO,NCTRY,NDTRY,NETRY,NFLAV,NGSPL,NSTRU,NSTRY,NZBIN,AZSOFT,
     & AZSPIN,CLDIR,HARDME,NOSPAC,PRNDEC,PRVTX,SOFTME,ZPRIME
C
      COMMON/HWPRCH/AUTPDF(2),BDECAY
C
C Parton shower common (same format as /HEPEVT/)
      PARAMETER (NMXPAR=500)
      COMMON/HWPART/NEVPAR,NPAR,ISTPAR(NMXPAR),IDPAR(NMXPAR),
     & JMOPAR(2,NMXPAR),JDAPAR(2,NMXPAR),PPAR(5,NMXPAR),VPAR(4,NMXPAR)
C
C Parton polarization common
      COMMON/HWPARP/DECPAR(2,NMXPAR),PHIPAR(2,NMXPAR),RHOPAR(2,NMXPAR),
     & TMPAR(NMXPAR)
C
C Electroweak boson common
      PARAMETER (MODMAX=5)
      COMMON/HWBOSC/ALPFAC,BRHIG(12),ENHANC(12),GAMMAX,RHOHEP(3,NMXHEP),
     & IOPHIG,MODBOS(MODMAX)
C
C Parton colour common
      COMMON/HWPARC/JCOPAR(4,NMXPAR)
C
C other HERWIG branching, event and hard subprocess common blocks
      COMMON/HWBRCH/ANOMSC(2,2),HARDST,PTINT(3,2),XFACT,INHAD,JNHAD,
     & NSPAC(7),ISLENT,BREIT,FROST,USECMF
C
      COMMON/HWEVNT/AVWGT,EVWGT,GAMWT,TLOUT,WBIGST,WGTMAX,WGTSUM,WSQSUM,
     & IDHW(NMXHEP),IERROR,ISTAT,LWEVT,MAXER,MAXPR,NOWGT,NRN(2),NUMER,
     & NUMERU,NWGTS,GENSOF
C
      COMMON/HWHARD/ASFIXD,CLQ(7,6),COSS,COSTH,CTMAX,DISF(13,2),EMLST,
     & EMMAX,EMMIN,EMPOW,EMSCA,EPOLN(3),GCOEF(7),GPOLN,OMEGA0,PHOMAS,
     & PPOLN(3),PTMAX,PTMIN,PTPOW,Q2MAX,Q2MIN,Q2POW,Q2WWMN,Q2WWMX,QLIM,
     & SINS,THMAX,TMNISR,TQWT,XX(2),XLMIN,XXMIN,YBMAX,YBMIN,YJMAX,
     & YJMIN,YWWMAX,YWWMIN,WHMIN,ZJMAX,ZMXISR,IAPHIG,IBRN(2),IBSH,
     & ICO(10),IDCMF,IDN(10),IFLMAX,IFLMIN,IHPRO,IPRO,MAPQ(6),MAXFL,
     & BGSHAT,COLISR,FSTEVT,FSTWGT,GENEV,HVFCEN,TPOL
C
C Arrays for particle properties (NMXRES = max no of particles defined)
      PARAMETER(NMXRES=400)
      COMMON/HWPROP/RLTIM(0:NMXRES),RMASS(0:NMXRES),RSPIN(0:NMXRES),
     & ICHRG(0:NMXRES),IDPDG(0:NMXRES),IFLAV(0:NMXRES),NRES,
     & VTOCDK(0:NMXRES),VTORDK(0:NMXRES)
C
      COMMON/HWUNAM/RNAME(0:NMXRES)
C
C Arrays for particle decays (NMXDKS = max total no of decays,
C                             NMXMOD = max no of modes for a particle)
      PARAMETER(NMXDKS=4000,NMXMOD=200)
      COMMON/HWUPDT/BRFRAC(NMXDKS),CMMOM(NMXDKS),DKLTM(NMXRES),
     & IDK(NMXDKS),IDKPRD(5,NMXDKS),LNEXT(NMXDKS),LSTRT(NMXRES),NDKYS,
     & NME(NMXDKS),NMODES(NMXRES),NPRODS(NMXDKS),DKPSET,RSTAB(0:NMXRES)
C
C Weights used in cluster decays
      COMMON/HWUWTS/REPWT(0:3,0:4,0:4),SNGWT,DECWT,QWT(3),PWT(12),
     & SWTEF(NMXRES)
C
C Parameters for cluster decays (NMXCDK = max total no of cluster
C                                         decay channels)
      PARAMETER(NMXCDK=4000)
      COMMON/HWUCLU/CLDKWT(NMXCDK),CTHRPW(12,12),PRECO,RESN(12,12),
     & RMIN(12,12),LOCN(12,12),NCLDK(NMXCDK),CLRECO
C
C Variables controling mixing and vertex information
      COMMON/HWDIST/EXAG,GEV2MM,HBAR,PLTCUT,VMIN2,VTXPIP(4),XMIX(2),
     & XMRCT(2),YMIX(2),YMRCT(2),IOPDKL,MAXDKL,MIXING,PIPSMR
C
C Arrays for temporarily storing heavy-b,c-hadrons decaying partonicaly
C (NMXBDK = max no such b-hadron decays in an event)
      PARAMETER (NMXQDK=20)
      COMMON/HWQDKS/VTXQDK(4,NMXQDK),IMQDK(NMXQDK),LOCQ(NMXQDK),NQDK
C
C Parameters for Sudakov form factors
C (NMXSUD= max no of entries in lookup table)
      PARAMETER (NMXSUD=1024)
      COMMON/HWUSUD/ACCUR,QEV(NMXSUD,6),SUD(NMXSUD,6),INTER,NQEV,NSUD,
     & SUDORD
C
      PARAMETER (NMXJET=200)
------------------------------------------------------------------------

                   ****** 8. FORM FACTOR FILE ******

   HERWIG uses look-up tables of Sudakov form factors for the evolution
   of initial-  and final-state parton showers.  These can be read from
   an input file  rather than being recomputed each time.  The reading,
   writing and computing of form factor tables is controlled by integer
   parameters LRSUD and LWSUD:

      LRSUD = N>0   Read form factors for this run from unit N
      LRSUD = 0     Compute new form factor tables for this run
      LRSUD < 0     Form factor tables are already loaded
      LWSUD = N>0   Write form factors on unit N for future use
      LWSUD = 0     Do not write new form factor tables

   The option LRSUD<0 allows the program to be initialized several times
   in the same run (e.g. to generate various event types) without recom-
   puting or rereading form factors.

   N.B. The Sudakov form factors depend on the parameters QCDLAM, VQCUT,
   VGCUT, NCOLO, NFLAV, NAFLA, RMASS(13) and RMASS(i) for i=1,...,NFLAV.
   Consequently form factor tables  MUST be recomputed every time any of
   these  parameters  is  changed.   From  version  5.1  onwards,  these
   parameters are written/read  with the  form factor tables  and checks
   are performed to ensure consistency.

   The parton  showering  algorithm  uses  the  two-loop  alpha_s,  with
   matching at each flavour threshold. However, the Sudakov table can be
   computed with either the one-loop or two-loop form,  according to the
   variable SUDORD  (= 1 or 2 respectively, DEFAULT=1).  If SUDORD=1 the
   two-loop value is  recovered using the  veto algorithm in the shower,
   whereas if SUDORD=2 no vetoes  are used in the final-state evolution.
   This means  that the relative weight of any shower  configuration can
   be calculated in a closed form,  hence that showers  can be `forced'.

   To next-to-leading order  the two  possibilities should be identical,
   but they  differ at beyond-NLO,  so some results may change a little.
   The most noticeable difference is that the form factor  table takes a
   factor of about five times longer to compute with SUDORD=2 than 1.
------------------------------------------------------------------------

                      ****** 9. EVENT DATA ******

   /HEPEVT/ is the standard common block containing current event data:

   NEVHEP      - event number
   NHEP        - number of entries for this event
   ISTHEP(I)   - status of entry I (see below)
   IDHEP(I)    - identity of entry I (revised Particle Data Group code)
   JMOHEP(1,I) - pointer to  first mother of entry I (see below)
   JMOHEP(2,I) - pointer to second mother of entry I (see below)
   JDAHEP(1,I) - pointer to first daughter of entry I (see below)
   JDAHEP(2,I) - pointer to  last daughter of entry I (see below)
   PHEP(*,I)   - (Px,Py,Pz,E,M) of entry I: M=sign(sqrt(abs(m**2)),m**2)
   VHEP(*,I)   - (x,y,z,t) of prod'n vertex of entry I (see section 13)

   All momenta are  given in the  laboratory frame,  in which  the input
   beam momenta are PBEAM1 and PBEAM2 as specified by the user and point
   along the +z and -z directions respectively.  Final  state  particles
   have ISTHEP(I) = 1.  See the next section for a  complete list of the
   special status codes used by HERWIG.

   The identity codes IDHEP are as those suggested by the LEP II Working
   group i.e. the revised Particle Data Group numbers plus the following

  * IDHEP = 91 for clusters, 94 for jets, 0 for others with no PDG code.

   (HERWIG also has its own internal identity codes  IDHW(I),  stored in
   /HWEVNT/. The utility subroutine HWUIDT translates between HERWIG and
   PDG identity codes.  See section 20 for further details.)

   The  mother/daughter  pointers are standard,  except that JMOHEP(2,I)
   and  JDAHEP(2,I) for a  PARTON are its  COLOUR  mother  and daughter,
   i.e., the partons to which its  colour and  anticolour are connected,
   respectively.  For this purpose the  primary partons from a hard sub-
   process are all  regarded as outgoing  (see examples in sects. 15, 19
   and 21).  Since quarks  have no  anticolour,  JDAHEP(2,I)  is used to
   point to  its FLAVOUR partner.  Similarly for JMOHEP(2,I) in the case
   of an antiquark.

   In addition to entries representing partons, particles, clusters etc,
   /HEPEVT/ contains purely informational entries representing the total
   c.m. momentum, hard and soft subprocess momenta, etc.  See section 10
   for the corresponding status codes.

   Information  from  all  stages  of event  processing  is  retained in
   /HEPEVT/ so the same particle may appear several times with different
   status codes. For example, an outgoing parton from a  hard scattering
   (entered initially with status 113 or 114) will appear after process-
   ing as an on-mass-shell parton before QCD branching (status 123,124),
   an off-mass-shell entry representing the flavour and momentum  of the
   outgoing  jet (status 143,144), and a jet constituent (157). It might
   also appear again in  other contexts,  e.g. as a spectator in a heavy
   flavour decay (status 154,160).

   Incoming partons  (entered with status  111, 112, changed to 121, 122
   after branching) give rise to spacelike jets (status 141,142, m**2<0,
   indicated by PHEP(5,IHEP)<0)  due to the loss of momentum via initial
   state  bremsstrahlung.  The  same  applies  in principle  to incoming
   leptons, but QED radiative corrections are not yet included.

   Each  parton jet  begins  with a  status 141-144 jet entry giving the
   total flavour and  momentum of the jet.  The first  mother pointer of
   this entry gives the  location of the  parent hard parton,  while the
   second gives that of the subprocess  c.m. momentum.  If QCD branching
   has occurred,  this is  followed by a  lightlike  CONE  entry,  which
   fixes the angular  extent of  the jet  and its  azimuthal orientation
   relative to the parton with which it interferes. The interfering par-
   ton is listed as the second mother of the cone.  Next come the actual
   constituents of the jet.  If no branching  has occurred,  there is no
   cone and the single jet constituent is the same as the jet.
------------------------------------------------------------------------

                    ****** 10. STATUS CODES ******

   A complete list of currently-used HERWIG status codes is given below.
   Many are used only in  intermediate stages of  event processing.  The
   most important for users are probably 1 (final-state particle), 101-3
   (initial  state), 141-4 (jets), and 199 (decayed  b-  and t-flavoured
   hadrons).

   The event status ISTAT in common /HWEVNT/ is roughly ISTHEP-100 where
   ISTHEP  is the status of entries being processed.  However, ISTAT=100
   for completed events.

                 +------+-------------------------------------------+
                 |ISTHEP|               Description                 |
                 +------+-------------------------------------------+
                 |   1  | final state particle                      |
                 |   2  | parton before hadronization               |
                 |   3  | documentation line                        |
                 +------+-------------------------------------------+
                 | 100  | cone limiting jet evolution               |
                 | 101  | `beam'   (beam 1)                         |
                 | 102  | `target' (beam 2)                         |
                 | 103  | overall centre of mass                    |
                 +------+-------------------------------------------+
                 | 110  | unprocessed hard process CoM              |
                 | 111  |      "      beam      parton              |
                 | 112  |      "      target       "                |
                 | 113  |      "      outgoing     "   3            |
                 | 114  |      "      outgoing     "   4            |
                 | 115  |      "      spectator    "                |
                 +------+-------------------------------------------+
                 |120-25| as 110-15, after processing               |
                 +------+-------------------------------------------+
                 | 130  | lepton in jet (unboosted)                 |
                 |131-34| as 141-44, unboosted to CoM               |
                 | 135  | spacelike parton (beam,   unboosted)      |
                 | 136  |     "        "   (target,     "    )      |
                 | 137  | spectator (beam,   unboosted)             |
                 | 138  |     "     (target,     "    )             |
                 | 139  | parton from branching   (unboosted)       |
                 | 140  |    "     "  g splitting (    "    )       |
                 +------+-------------------------------------------+
                 |141-44| jet from parton type 111-14               |
                 |145-50| as 135-40 boosted, unclustered            |
                 +------+-------------------------------------------+
                 | 151  | as 159, not yet clustered                 |
                 | 152  | as 160,  "   "      "                     |
                 | 153  | spectator from beam                       |
                 | 154  |     "       "  target                     |
                 | 155  | heavy quark before decay                  |
                 | 156  | spectator before heavy decay              |
                 | 157  | parton from QCD branching                 |
                 | 158  |   "    after gluon splitting              |
                 | 159  |   "    from cluster splitting             |
                 | 160  | spectator after heavy decay               |
                 +------+-------------------------------------------+
                 | 161  | beam   spectator after gluon splitting    |
                 | 162  | target     "       "     "       "        |
                 | 163  | other  cluster before soft process        |
                 | 164  | beam      "       "     "     "           |
                 | 165  | target    "       "     "     "           |
                 | 167  | unhadronized beam   cluster               |
                 | 168  | unhadronized target cluster               |
                 +------+-------------------------------------------+
                 | 170  | soft process centre of mass               |
                 | 171  | soft cluster (beam,   unhadronized)       |
                 | 172  | soft cluster (target,       "     )       |
                 | 173  | soft cluster (other,        "     )       |
                 +------+-------------------------------------------+
                 | 181  | beam         cluster (no soft process)    |
                 | 182  | target          "    ( "   "     "   )    |
                 | 183  | hard process    "    (hadronized)         |
                 | 184  | soft            "    (beam,   hadronized) |
                 | 185  |   "             "    (target,      "    ) |
                 | 186  |   "             "    (other,       "    ) |
                 +------+-------------------------------------------+
                 |190-93| as 195-98, before decays                  |
                 | 195  | direct unstable non-hadron                |
                 | 196  |   "        "    hadron (1-body cluster)   |
                 | 197  |   "        "       "   (2-body cluster)   |
                 | 198  | indirect unstable hadron or lepton        |
                 | 199  | decayed heavy flavour hadron              |
                 +------+-------------------------------------------+
                 | 200  | neutral B meson, flavour at production    |
                 +------+-------------------------------------------+
------------------------------------------------------------------------

                    ****** 11. EVENT WEIGHTS ******

   The  default is  to generate  unweighted events  (EVWGT=AVWGT).  Then
   event distributions are generated by computing a  weight proportional
   to the cross section  and comparing it with a random number times the
   maximum weight.  Set WGTMAX to the maximum weight, or to zero for the
   program to compute it.  If a weight greater than  WGTMAX is generated
   during execution, a warning is printed and WGTMAX is reset. Similarly
   if the  efficiency  is too low  (WGTMAX too large).  If these  errors
   occur  too often,  output  event  distributions  could be  distorted.

   To generate  weighted  events,  set NOWGT=.FALSE. in common /HWEVNT/.

   In QCD hard scattering and heavy flavour and direct photon production
   (IPROC = 1500 to 1800) the transverse energy distribution of weighted
   events (or the efficiency for unweighted  events) can be varied using
   the parameters PTMIN, PTMAX and PTPOW.

   Similarly in  Drell-Yan processes (IPROC = 13**) the lepton pair mass
   distribution is controlled by the parameters  EMMIN, EMMAX and EMPOW,
   and in  deep  inelastic  scattering the  Q**2 distribution  is set by
   Q2MIN, Q2MAX and Q2POW.

   Data on weights  generated are output at the end of the run. The mean
   weight is an  estimate of the cross section (in nanobarns) integrated
   over the region used for event generation.

   N.B.  The  mean weight  is the  sum of  weights  divided by the total
   number  of  WEIGHTS  generated,  not  the  total  number  of  EVENTS.
------------------------------------------------------------------------

                ****** 12. HEAVY FLAVOUR DECAYS ******

    Heavy quark decays are treated as secondary hard  subprocesses.  Top
    quarks can decay either before or after hadronization,  depending on
    the value of the logical variable  DECAY  returned by the subroutine
    HWDTOP.  At present decay occurs before hadronization (DECAY=.TRUE.)
    if the top mass is above 130 GeV (default=170 GeV). Any hypothetical
    heavier quarks  always decay before hadronization.  Top- and bottom-
    flavoured hadrons are split into collinear heavy quark and spectator
    and the former decays independently. After decay, parton showers may
    be  generated  from  coloured decay products, in the usual way.  See
    Nucl. Phys. B330 (1990) 261 for details  of the  treatment of colour
    coherence in these showers.

    The arrays  FBTM, FTOP & FHVY which were used in versions before 5.9
    to store the bottom, top & heavier quarks' partonic  decay fractions
    are gone.  Such decays are specified in the  decay tables like other
    particles' decay modes: this permits different decays to be given to
    individual heavy hadrons.  Changes to the decay table entries can be
    made on an event by event basis if desired. Partonic decays of charm
    hadrons and quarkonium states are also now supported.  The products'
    order in a  partonic decay mode is significant.  For example, if the
    decay is Q --> W+q --> (f+fbar')+q occurring inside a Q-sbar hadron,
    the required ordering is:

             Q+sbar --->(f+fbar')+(q+sbar)
                     or (q+fbar')+(f+sbar)  `colour rearranged'

    In both cases the (V-A)^2 ME^2 is proportional to: p_0.p_2 * p_1*p_3

    The structure of the program has been altered so that secondary hard
    subrocess and subsequent fragmentation associated with each partonic
    heavy hadron decay appear separately. Thus pre-hadronization t quark
    decays are treated individually  as are any subsequent bottom hadron
    partonic decays.

    Additionally decays of heavy hadrons to exclusive non-partonic final
    states are supported. No check against double counting from partonic
    modes is included. However this isn't expected to be a major problem
    for the semi-leptonic and 2-body hadronic modes supplied.
------------------------------------------------------------------------

           ****** 13. SPACE-TIME STRUCTURE OF EVENTS ******

    The space-time structure of events is now available for all types of
    subprocess. The production vertex of each: parton, cluster, unstable
    resonance and final state particle is supplied in the VHEP(4,NMXHEP)
    array of /HEPEVT/; set PRVTX=.TRUE. to include this information when
    printing the event record (120 column format).  The units are: x,y,z
    in mm and t mm/c. In the case of partons and clusters the production
    points are always given in a loacl coordinate system centered on the
    their hard sub-process. This helps seperate the fermi scale partonic
    showers from millimeter scale distances possible in particle decays,
    for example the partonic decays of heavy (c,b) hadrons. The vertices
    of hadrons produced in cluster decays are always corrected back into
    the laboratory coordinate system.

    It is possible to vary the principal interaction point,  assigned to
    the CMF (ISTHEP=103) track, by setting PIPSMR=.TRUE. The smearing is
    generated by the routine HWRPIP according to a triple Gaussian,  see
    the code for details.  Also,  it is possible to veto particle decays
    that would occur outside a specified volume by setting MAXDKL=.TRUE.
    Each putative decay is tested in HWDXLM and if it would have decayed
    outside the chosen volume  it is frozen and labelled as final state.
    Using IOPDKL = 1,2 selects a cylindrical or spherical allowed region
    (about the origin) see the code for details.

    Lepton and hadron lifetimes are supplied in the array RLTIM(NMXRES).
    The lifetimes of heavy quarks (TQRK, VQRK, AQRK, HQRK AND HPQK), and
    weak bosons (W+, W-, Z0/GAMA*, HIGGS and Z0P) are derived from their
    calculated or specified widths as calculated in HWUDKS, whilst light
    quarks and gluons are given an effective minimum width, sqrt(VMIN2),
    that acts as a lifetime cut-off - see below.  Recall that the proper
    lifetime = HBAR/Gamma. All particles whose lifetimes are larger than
    PLTCUT are set stable.

    The proper (= rest frame) time at which an unstable lepton or hadron
    decays is generated according to the exponential decay law with mean
    lifetime <tau>=RLTIM. The laboratory frame decay time  and  distance
    travelled are obtained by applying a boost:

    Rest    Prob (proper time < t) =  1   * exp(-t/<tau>)
    frame                           <tau>  

    Lab.    time =        gamma * proper time    beta = v/c
    frame   dist = beta * gamma * proper time   gamma = 1/sqrt(1-beta^2)

    The production vertices of  the daughter particles are calculated by
    adding the distance travelled by the mother particle  as given above
    to its production vertex.  A similar prescription is used for parton
    showers: proper lifetimes are taken from an exponential distribution
    with a virtuality dependent mean lifetime 1/HBAR*sqrt(q^2/(q^2-m^2))
    inspired by the uncertainty relationship: mean lifetimes are limited
    by a cut-off on the minimum virtuality VMIN2.  The mean lifetimes of
    heavy quarks and weak bosons, which can have appreciable widths, are
    given by:

                                    hbar.sqrt(q^2)
              <tau>(q^2) =   ----------------------------- 
                           \/(q^2-M^2)^2 + (Gamma.q^2/M)^2

    As this formula has the appropriate limits for vanishing virtuality,
    q^2=m^2, or width, gamma=0, it is actually also used in the hadronic
    and partonic showers: see HWUDKL.

    In the case of cluster the initial production vertex is taken as the
    midpoint of a line perpendicular to the cluster's direction and with
    pair. If such a cluster undergoes a forced splitting to two clusters
    the string picture is adopted. The vertex of the light quark pair is
    positioned so that the masses of the two daughter clusters  would be
    the same as that for two equivalent string fragments. The production
    vertices of the daughter clusters are given by the first crossing of
    their constituent q-qbar pairs.  This part of the space-time picture
    is admittedly ad hoc however no physics depends upon it.

    When MIXING=.TRUE. particle - antiparticle mixing for B^0_d,s mesons
    is implimented. The probability that a meson is mixed when it decays
    is given in terms of its lab-frame decay time by:

                1     sin(X*m*t/c<tau>E)      X=Delta-M    Y=Delta-Gamma
    Prob(mix) = - + ----------------------      -------      -----------
                2   2 *cosh(Y*m*t/c<tau>E)       Gamma        2 * Gamma

    The ratios X and Y are stored in XMIX(I) & YMIX(I), I=1,2 for q=s,d.
    Whenever a neutral B meson occurs in an event a copy of the original
    track is always added to the event record, with ISTHEP=200, it gives
    the particle's flavour at the production (cluster decay) time.  This
    is in addition to the usual decaying particle, ISTHEP=19*, track.
------------------------------------------------------------------------

             ****** 14. COLOUR REARRANGEMENT MODEL ******

    HERWIG now contains a colour rearrangement model based on the space-
    time structure of an event occuring at the end of the parton shower.
    This is illustrated in the simple example shown below where a colour
    neutral source results in a q-g-g-qbar  shower.  In the conventional
    hadronization model after a nonperturbative splitting of final state
    gluons  - Wolfram ansatze -  colour singlet clusters are formed from
    neighbouring q-qbar pairs: (ij)(pq)(kl).  However when CLRECO=.TRUE.
    the program first creates colour singlet clusters as normal but then
    checks all (non-neighbouring) pairs of clusters  to test if a colour
    rearrangement lowers the sum of the clusters' spatial sizes added in
    quadrature. A cluster's size is defined to be the Lorentz invariant,
    space-time distance between the constituent  quark's and anitquark's
    production points. If an allowed alternative is found, that is:

    (ij)(kl) --> (il)(jk) s.t. (|d_ij|^2+|d_kl|^2) > (|d_il|^2+|d_kl|^2)

    then it is accepted with a probability given by PRECO (default 1/9).

                  ____ i     Normal:           (ij) (pq) (kl)
                 /
                /____/ j     If:
               ------
              /      \ p        |d_ij|^2+|d_kl|^2 > |d_il|^2+|d_kl|^2
       ------|
              \______/ q     colour rearr.:    (il) (pq) (jk)
                -----
                \    \ k     Not allowed:      (iq) (jp) (kl)
                 \                                   ^
                  ---- l                             | colour octet

    Note that not all colour rearrangements are allowed, for instance in
    the example (ij)(pq) --> (ip)(jq) the cluster (jq) is a colour octet
    - it contains both products from a non-perturbative gluon splitting.

    Multiple colour rearrangements are considered by the program, as are
    those between clusters in jets arising from a single, colour neutral
    source - for example Z0 decay (as shown above) - or due to more than
    one source - for example e+e- --> W+W- --> 4 jets. In the later case
    a new parameter, EXAG, is available to artificially scale the W - or
    other weak boson - lifetimes so that any dependence of rearrangement
    effects on source separation can be investigated. The CLRECO  option
    can be used for all the processes available in HERWIG.

    ** NOTE **  Before using the program with CLRECO=.TRUE. for detailed
    physics analyses the default parameters should be retuned to  `lower
    energy' data with this option switched on.
------------------------------------------------------------------------

                ****** 15. QCD HARD SUBPROCESSES ******

    At present only 2->2 subprocesses are implemented. They are class-
    ified as shown below.

            +-----+------------------------------+---------+
            |IHPRO|   Process 1 + 2 -> 3 + 4     |Col/F.Con|
            +-----+------------------------------+---------+
            |  1  | q + q        -> q + q        | 3 4 2 1 |
            |  2  | q + q        -> q + q        | 4 3 1 2 |
            |  3  | q + q'       -> q + q'       | 3 4 2 1 |
            |  4  | q + qbar     -> q'+ qbar'    | 2 4 1 3 |
            |  5  | q + qbar     -> q + qbar     | 3 1 4 2 |
            |  6  | q + qbar     -> q + qbar     | 2 4 1 3 |
            |  7  | q + qbar     -> g + g        | 2 4 1 3 |
            |  8  | q + qbar     -> g + g        | 2 3 4 1 |
            |  9  | q + qbar'    -> q + qbar'    | 3 1 4 2 |
            | 10  | q + g        -> q + g        | 3 1 4 2 |
            | 11  | q + g        -> q + g        | 3 4 2 1 |
            | 12  | qbar + q     -> qbar' +q'    | 3 1 4 2 |
            | 13  | qbar + q     -> qbar + q     | 2 4 1 3 |
            | 14  | qbar + q     -> qbar + q     | 3 1 4 2 |
            | 15  | qbar + q     -> g + g        | 3 1 4 2 |
            | 16  | qbar + q     -> g + g        | 4 1 2 3 |
            | 17  | qbar + q'    -> qbar + q'    | 2 4 1 3 |
            | 18  | qbar + qbar  -> qbar + qbar  | 4 3 1 2 |
            | 19  | qbar + qbar  -> qbar + qbar  | 3 4 2 1 |
            | 20  | qbar + qbar' -> qbar + qbar' | 4 3 1 2 |
            | 21  | qbar + g     -> qbar + g     | 2 4 1 3 |
            | 22  | qbar + g     -> qbar + g     | 4 3 1 2 |
            | 23  | g + q        -> g + q        | 2 4 1 3 |
            | 24  | g + q        -> g + q        | 3 4 2 1 |
            | 25  | g + qbar     -> g + qbar     | 3 1 4 2 |
            | 26  | g + qbar     -> g + qbar     | 4 3 1 2 |
            | 27  | g + g        -> q + qbar     | 2 4 1 3 |
            | 28  | g + g        -> q + qbar     | 4 1 2 3 |
            | 29  | g + g        -> g + g        | 4 1 2 3 |
            | 30  | g + g        -> g + g        | 4 3 1 2 |
            | 31  | g + g        -> g + g        | 2 4 1 3 |
            +-----+------------------------------+---------+

   `Col/F.Con'  refers  to the  colour/flavour  connections  between the
   partons:`I J K L' means that the colour of parton 1 comes from parton
   I, that of 2 from J, etc.  For antiquarks, which have no colour (only
   anticolour),  the label shows  instead to which parton the flavour is
   connected.  For this colour/flavour labelling all partons are defined
   as outgoing.  Thus, for example,  process  10 has colour  connections
   3 1 4 2, corresponding to the colour flow diagram:

                           1 -->--+ +-->-- 3
                                  | |
                                  | |
                             --<--+ +--<--
                           2 -->------->-- 4

   When different colour flows are possible, they are listed as separate
   subprocesses.  This separation  is not exact  but is  normally a good
   approximation.  The sum of the colour flows is the exact lowest-order
   cross section.
------------------------------------------------------------------------

           ****** 16. QCD DIRECT PHOTON SUBPROCESSES ******

            +-----+------------------------------+---------+
            |IHPRO|   Process 1 + 2 -> 3 + 4     |Col/F.Con|
            +-----+------------------------------+---------+
            | 41  | q + qbar     -> g + photon   | 2 3 1 4 |
            | 42  | q + gluon    -> q + photon   | 3 1 2 4 |
            | 43  | qbar + q     -> g + photon   | 3 1 2 4 |
            | 44  | qbar + gluon -> qbar + photon| 2 3 1 4 |
            | 45  | gluon + q    -> q    + photon| 2 3 1 4 |
            | 46  | gluon + qbar -> qbar + photon| 3 1 2 4 |
            | 47  | gluon + gluon-> gluon+ photon| 2 3 1 4 |
            +-----+------------------------------+---------+
            | 51  | photon+ q    -> gluon+ q     | 1 4 2 3 |
            | 52  | photon+ qbar -> gluon+ qbar  | 1 3 4 2 |
            | 53  | photon+ gluon-> q    + qbar  | 1 4 2 3 |
            +-----+------------------------------+---------+
            | 61  | q + qbar     -> photon+photon| 2 1 3 4 |
            | 62  | qbar + q     -> photon+photon| 2 1 3 4 |
            | 63  | gluon + gluon-> photon+photon| 2 1 3 4 |
            +-----+------------------------------+---------+
            | 71  | photon+ q    -> M(S=0) +q'   | 1 4 3 2 |
            | 72  | photon+ q    -> M(S=1)L+q'   | 1 4 3 2 |
            | 73  | photon+ q    -> M(S=1)T+q'   | 1 4 3 2 |
            | 74  | photon+ qbar -> M(S=0) +qbar'| 1 4 3 2 |
            | 75  | photon+ qbar -> M(S=1)L+qbar'| 1 4 3 2 |
            | 76  | photon+ qbar -> M(S=1)T+qbar'| 1 4 3 2 |
            +-----+------------------------------+---------+

   N.B. The photon is connected to itself.
------------------------------------------------------------------------

           ****** 17. QCD HIGGS PLUS JET SUBPROCESSES ******

            +-----+------------------------------+---------+
            |IHPRO|   Process 1 + 2 -> 3 + 4     |Col/F.Con|
            +-----+------------------------------+---------+
            | 81  | q    + qbar  -> g    + H     | 2 3 1 4 |
            | 82  | q    + g     -> q    + H     | 3 1 2 4 |
            | 83  | qbar + q     -> g    + H     | 3 1 2 4 |
            | 84  | qbar + g     -> qbar + H     | 2 3 1 4 |
            | 85  | g    + q     -> q    + H     | 2 3 1 4 |
            | 86  | g    + qbar  -> qbar + H     | 3 1 2 4 |
            | 87  | g    + g     -> g    + H     | 2 3 1 4 |
            +-----+------------------------------+---------+

   N.B. The Higgs is connected to itself.
------------------------------------------------------------------------

               ****** 18. ELECTROWEAK SUBPROCESSES ******

   HERWIG  generates  Higgs bosons  through  gluon-gluon/quark-antiquark
   fusion,  and W fusion in hadron-hadron collisions  (IPROC=1600+ID and
   1900+ID),  in lepton-lepton  collisions  through  the Bjorken process
   (that is, Z(*)->Z(*)H  with one or both Zs  off-shell)  and  W fusion
   (IPROC=300+ID and 400+ID),  and in lepton-hadron collisions through W
   fusion (IPROC=9500+ID).  Each process  is generated  according to the
   exact  leading  order matrix  element in the s-channel approximation.
   This results in  unitarity violation for  Mh >> Mw,  s >~ a few Mh^2,
   (where s=qh^2), so to regularize this,  the Mh*GAMH in the propagator
   can be replaced by SQRT(s)*GAMH(s). The variable IOPHIG controls this
   procedure:

       +------+------------------------------+-----------+
       |IOPHIG|   Choose s according to      | Reweight? |
       +------+------------------------------+-----------+
       |   0  | s^2 / ((s-Mh^2)^2 + Mh*GAMH) |    YES    |
       |   1  |  1  / ((s-Mh^2)^2 + Mh*GAMH) |    YES    |
       |   2  | s^2 / ((s-Mh^2)^2 + Mh*GAMH) |     NO    |
       |   3  |  1  / ((s-Mh^2)^2 + Mh*GAMH) |     NO    |
       +------+------------------------------+-----------+

   Where reweighting means weighting the distribution back to

                          SQRT(s) * GAMH(s)
                     ----------------------------
                     (s-Mh^2)^2 + SQRT(s)*GAMH(s)

   The default is IOPHIG=1.  The difference  between options  0 and 1 is
   purely  in the  weight  distribution  produced.  Options  2 and 3 are
   intended primarily for  users who wish to supply  their own unitarity
   conserving  reweighting  function at the point  indicated in  routine
   HWHIGM.  In all cases,  the distribution  is restricted  to the range
   [Mh-GAMMAX*GAMH , Mh+GAMMAX*GAMH].  GAMMAX defaults to 10, but in the
   (probably unphysical) region  Mh >~ 1TeV should be reduced to protect
   against poor weight distributions. These considerations do not affect
   the distribution noticably for  Mh <~ 500 GeV,  and GAMMAX can safely
   be increased if necessary.

   For each process, ID controls the Higgs decay: ID=1-6 for quarks, 7-9
   for leptons, 10/11 for WW/ZZ pairs,  and 12 for photons.  In addition
   ID=0 gives quarks of all flavours,  and ID=99  gives all decays.  For
   each process,  the average  event  weight  is the cross section in nb
   times the branching  fraction  to the requested decay.  The branching
   ratios to quarks use the next-to-leading logarithm corrections, those
   to WW/ZZ pairs allow for one or both bosons off-shell. The amplitudes
   for all Higgs vertices are multiplied by the factor ENHANC(ID)  where
   ID is the same as in IPROC=300+ID except the gammagammaHiggs `vertex'
   which is calculated from ENHANC(6)  and ENHANC(10)  for the top and W
   loops.  This allows  the simulation  of any chargeless  scalar Higgs.
   Note however  that  pseudoscalar and charged Higgses,  and  processes
   involving more than one Higgs (eg the decay H-->hZ) are not included.

   Gauge bosons  are generated  through the  processes  of  W + 1 parton
   production in hadron-hadron  collisions,  and  WW  pair production in
   lepton-lepton collisions, as well as in the Higgs processes mentioned
   above.   In  all  cases  their  decay is controlled  by the  variable
   MODBOS(i). This controls the decay of the ith gauge boson per event:

       +---------+-----------------+-----------------+
       |MODBOS(i)|     W Decay     |     Z Decay     |
       +---------+-----------------+-----------------+
       |    0    |        all      |        all      |
       |    1    |       qqbar     |       qqbar     |
       |    2    |        enu      |       e+e-      |
       |    3    |        munu     |      mu+mu-     |
       |    4    |       taunu     |     tau+tau-    |
       |    5    |     enu & munu  |    ee & mumu    |
       |    6    |        all      |      nunu       |
       |    7    |        all      |     bbbar       |
       |   >7    |        all      |       all       |
       +---------+-----------------+-----------------+

   All  entries of  MODBOS  default to 0.   Bosons which are produced in
   pairs  (ie. from WW pair production, or Higgs decay)  are symmetrized
   in MODBOS(i)  and  MODBOS(i+1).  For processes which directly produce
   gauge bosons, the event weight includes the branching fraction to the
   requested decay,  but this is only true for Higgs production if decay
   to WW/ZZ is forced (ID=10/11) and not if ID=99. The spin-correlations
   in the decays are handled in one of two ways:
   (a) the diagonal  members of the  spin  density  matrix are stored in
       RHOHEP(i,IHEP),  where i=1,2,3 for helicity=i-2 in the centre-of-
       mass frame of their production,  for processes  where this matrix
       is diagonal (ie. there is no interference between spin states).
   (b) the  correlations  in the  decay  are  handled  directly  by  the
       production routine where (a) is not possible.
   In the case of  gamma gamma --> W W  the  decay correlations  are not
   correctly included: they currently decay isotropically.

   The electroweak vector boson--fermion coupling  constants are  stored
   in the arrays QFCH(I), VFCH(I,J) and AFCH(I,J) for the charge, vector
   and axial vector couplings to the neutral current respectively. These
   are given in the convention 
       V_f=(T_3/2-Qsin^2_W)/(cos_W sin_W);  A_f=T_3/(2 cos_W sin_W).
   In each case,
     I= 1- 6: d,u,s,c,b,t (quarks)
      =11-16: e,nu_e,mu,nu_mu,tau,nu_tau (leptons) (`I=IDHW-110')
     J=1 for minimal SM:
      =2 for Z' couplings (only included if ZPRIME=.TRUE.)
   Note that no universality is assumed -- couplings can be arbitrarily
   set for each fermion species separately.
   The quark mixing matrix  is stored in VCKM(K,L),  K=1,2,3 for u,c,t,
   L=1,2,3 for d,s,b.

   A running electromagnetic coupling constant is provided, HWUAEM(Q2).
   ALPHEM =1/137 provides the normalisation at the Thomson (Q2=0) limit
   and is used for all processes involving real photons.
   The electroweak coupling is calculated as,
                 g^2 = 4 PIFAC ALPHEM(Q2) / SWEIN,
   where Q2 is appropriate for the given process.
   Photon emission in parton showers,  and in the  `dead-zone'  in e+e-
   is enhanced by a factor of ALPFAC (default=1.).
------------------------------------------------------------------------

             ****** 19. INCLUDING NEW SUBPROCESSES ******

   It should not be difficult for users to include  further subprocesses
   in this version of the program if required.  The parton and hard sub-
   process 4-momenta,  masses and  identity codes  need to be entered in
   COMMON/HEPEVT/ with the appropriate status codes ISTHEP(I)=110-114 to
   tell the program which is which (see table in sect. 10).  The colour/
   flavour structure should be specified by the second mother and daugh-
   ter pointers as explained in section 9   (see also the sample output
   and guide, sections. 20 and 21).

   Apart from the status codes ISTHEP, the HERWIG identity codes IDHW(I)
   in COMMON/HWEVNT/ also need  to be set correctly.  The IDHW codes can
   be listed in a run with  IPRINT=2:  the most important are the quarks
   1-6 (as IDHEP), antiquarks 7-12, gluon 13, overall c.m. 14, hard c.m.
   15, soft c.m. 16,  photon 59,  leptons 121-126,  antileptons 127-132.

   The  utility  subroutine  HWUIDT(IOPT,IPDG,IHWG,NAME)  is provided to
   translate between  Particle Data Group code  IPDG,  HERWIG code IHWG,
   and HERWIG  character*8 NAME,  with IOPT=1,2,3  depending on which of
   IPDG, IHWG and NAME is the input argument.

   Consider for example  the process of  virtual photon-gluon  fusion to
   make  b+bbar in  e p collisions.

       **** N.B. This process is now included as IPROC = 9102 ****

   We assume the user provides a subroutine to generate the momenta PHEP
   for the  hard  subprocess   e+g -> e+b+bbar.  The colour structure is

                        (e)4 ........... 7(e)
                                  :
                                  :
                                  +-->-- 8(b)
                                  |
                             -->--+
                        (g)5 --<-----<-- 9(bbar)

   Thus the momenta generated, together with those of the initial beams
   and the overall  centre of mass,  could be entered  in the following
   sequence:

             +----+--------+------+-----+------+------+----+
             |IHEP|  Entry |ISTHEP|IDHEP|JMOHEP|JDAHEP|IDHW|
             +----+--------+------+-----+------+------+----+
             |  1 | e beam |  101 |   11|  0  0|  0  0| 121|
             |  2 | p beam |  102 | 2212|  0  0|  0  0|  73|
             |  3 | ep c.m.|  103 |    0|  0  0|  0  0|  14|
             +----+--------+------+-----+------+------+----+
             |  4 | e in   |  111 |   11|  6  7|  0  7| 121|
             |  5 | gluon  |  112 |   21|  6  9|  0  8|  13|
             |  6 | hard cm|  110 |    0|  4  5|  7  9|  15|
             |  7 | e out  |  113 |   11|  6  4|  0  4| 121|
             |  8 | b      |  114 |    5|  6  5|  0  9|   5|
             |  9 | bbar   |  114 |   -5|  6  8|  0  5|  11|
             +----+--------+------+-----+------+------+----+

   Note that if there are more than two outgoing partons,  the first has
   status  113 and all the others 114.  Each parton has JMOHEP(1,I)=6 to
   indicate  the location of  the hard c.m.  for this subprocess,  while
   JMOHEP(2,I) gives the location of the colour mother (treating the in-
   coming gluon as outgoing) or the connected electron. JDAHEP(1,I) will
   be set by the jet generator  HWBGEN,  while JDAHEP(2,I) points to the
   anticolour mother (or connected electron). Finally the HERWIG identi-
   fiers IDHW(I) could be set to the indicated values by means of the
   translation subroutine HWUIDT as follows:

      CHARACTER*8 NAME
      .....
      NHEP=9
      IDHEP(1)=11
      IDHEP(2)=2212
      .....
      IDHEP(9)=-5
      DO 10 I=1,NHEP
   10 CALL HWUIDT(1,IDHEP(I),IDHW(I),NAME)
      IDHW(6)=15

   The last statement is needed because  IDPDG(I)=0  returns IDHW(I)=14.
   If subroutine HWBGEN is now called, it will find the coloured partons
   and generate QCD jets from them.  Subsequent calls to  HWCFOR etc can
   then be used to form clusters and hadronize them.

   If the  hard subprocess  routine is  called from  HWEPRO,  like those
   already provided, it should have two options controlled by the logic-
   al variable  GENEV  in  COMMON/HWHARD/.   For GENEV=.FALSE., an event
   weight  (normally the  cross section  in nanobarns)  is generated and
   stored as  EVWGT  in COMMON/HWEVNT/.  If this  weight is  accepted by
   HWEPRO, the subroutine is called a second time with  GENEV=.TRUE. and
   the corresponding  event data should  then be generated and stored as
   explained above.
------------------------------------------------------------------------

                  ****** 20. ERROR CONDITIONS ******

   Certain combinations of input parameters may lead to problems in exe-
   cution.  HERWIG  tries to  detect these and  print a warning.  Errors
   during  execution are dealt with by  HWWARN  which prints the calling
   subprogram and a code and takes  appropriate action.  In general, the
   larger  the code  the more serious  the problem.  Refer to the source
   code  to find  out why  HWWARN  was  called.  Events  can be rerun by
   setting the random number seeds  NRN to the values given in the error
   message or event dump,  and MAXWGT to the maximum weight  encountered
   in the run.  Contents of  /HEPEVT/  can by printed by calling HWUEPR,
   those of /HWPART/ (last parton shower) by HWUBPR.

   If WGTMAX is increased during event generation, so that this message
   is printed:
   HWWARN CALLED FROM SUBPROGRAM HWEPRO: CODE =   1
   EVENT      21:   SEEDS =  836291635 & 1823648329  WEIGHT = 0.3893E-08
   EVENT SURVIVES. EXECUTION CONTINUES
          NEW MAXIMUM WEIGHT = 0.428217360829367E-08
   then to regenerate any later events, WGTMAX must be set to the printed
   value, as well as setting NRN to the appropriate seeds.

   Examples of error messages:

   HWWARN CALLED FROM SUBPROGRAM HWSBRN: CODE = 101
   EVENT      31:   SEEDS =  422399901 &  771980111  WEIGHT = 0.3893E-08
   EVENT KILLED.   EXECUTION CONTINUES

   Spacelike (initial-state) parton branching had no phase space. This
   can happen due to cutoffs which are slightly different in the hard
   subprocess and the parton shower.
   Action taken:  program  throws away  this event and starts a new one.

   HWWARN CALLED FROM SUBPROGRAM HWCHAD: CODE = 102
   EVENT      51:   SEEDS = 1033784787 & 1428957533  WEIGHT = 0.3893E-08
   EVENT KILLED.   EXECUTION CONTINUES

   A cluster has been formed with too low a mass to represent any hadron
   of the correct flavour, and there is no colour-connected cluster from
   which the necessary additional mass could be transferred.
   Action taken:  program  throws away  this event and starts a new one.

   HWWARN CALLED FROM SUBPROGRAM HWUINE: CODE= 200
   EVENT SURVIVES.  RUN ENDS GRACEFULLY

   CPU time limit liable to be reached  before generating  MAXEV events.
   Action taken:  skips to terminal  calculations using existing events.

   HWWARN CALLED FROM SUBPROGRAM HWBSUD: CODE= 500
   RUN CANNOT CONTINUE

   The table of Sudakov form factors read on unit  LRSUD does not extend
   to  the  maximum  momentum  scale  (QLIM)  specified  for  this  run.
   Action taken: run aborted.  The user must either reduce  QLIM  or set
   LRSUD=0  to make a  bigger  table  (set  LWSUD  nonzero to write it).

   HWWARN CALLED FROM SUBPROGRAM HWBSUD: CODE= 515
   RUN CANNOT CONTINUE

   The table of  Sudakov form factors read on unit  LRSUD is for a diff-
   erent value of a relevant parameter  (in this case the b quark mass).
   Action taken: run aborted.  The user must make a new table (set LWSUD
   nonzero to write it).
------------------------------------------------------------------------
                    ****** 21. SAMPLE OUTPUT ******

   Below we give a  complete listing  of output  from version 5.9 of the
   program,  set up for  t quark  production in  pbar-p  collisions at a
   c.m. energy of  1.8 TeV.  To shorten the event record, the underlying
   event has been turned off (IPROC = 11706) and production vertices are
   not  printed  (PRVTX=.FALSE.).   The  main features of the output are
   discussed in section 22.

         HERWIG 5.9    22nd July 1996

         Please reference: G. Marchesini, B.R. Webber,
         G.Abbiendi, I.G.Knowles, M.H.Seymour & L.Stanco
         Computer Physics Communications 67 (1992) 465

         INPUT CONDITIONS FOR THIS RUN

         BEAM 1 (PBAR    ) MOM. =    900.00
         BEAM 2 (P       ) MOM. =    900.00
         PROCESS CODE (IPROC)   =   11706
         NUMBER OF FLAVOURS     =    6
         STRUCTURE FUNCTION SET =    5
         AZIM SPIN CORRELATIONS =    T
         AZIM SOFT CORRELATIONS =    T
         QCD LAMBDA (GEV)       =    0.1800
         DOWN     QUARK  MASS   =    0.3200
         UP       QUARK  MASS   =    0.3200
         STRANGE  QUARK  MASS   =    0.5000
         CHARMED  QUARK  MASS   =    1.5500
         BOTTOM   QUARK  MASS   =    4.9500
         TOP      QUARK  MASS   =  170.0000
         GLUON EFFECTIVE MASS   =    0.7500
         EXTRA SHOWER CUTOFF (Q)=    0.4800
         EXTRA SHOWER CUTOFF (G)=    0.1000
         PHOTON SHOWER CUTOFF   =    0.4000
         CLUSTER MASS PARAMETER =    3.3500
         SPACELIKE EVOLN CUTOFF =    2.5000
         INTRINSIC P-TRAN (RMS) =    0.0000
         MIN P-TRAN FOR 2->2    =   10.0000
         MAX P-TRAN FOR 2->2    =  900.0002

         NO EVENTS WILL BE WRITTEN TO DISK

         B_d: Delt-M/Gam =0.7000 Delt-Gam/2*Gam =0.0000
         B_s: Delt-M/Gam = 10.00 Delt-Gam/2*Gam =0.2000

         PDFLIB NOT USED FOR BEAM 1
         PDFLIB NOT USED FOR BEAM 2

         Checking consistency of particle properties

         Checking consistency of decay tables

Line,  565 decay: LMBDA_C+ --> XI*0     K*+                                
is kinematically not allowed, Min-Mout=     -0.139
LMBDA_C+: BR sum = 0.97800
Rescaling to 1

Line,  990 decay: LMBDA_C- --> XI*BAR   K*-                                
is kinematically not allowed, Min-Mout=     -0.139
LMBDA_C-: BR sum = 0.97800
Rescaling to 1

         PARTICLE TYPE  21=PI0      SET STABLE

         INITIAL SEARCH FOR MAX WEIGHT

         PROCESS CODE IPROC =       11706
         RANDOM NO. SEED 1  =     1246579
                    SEED 2  =     8447766
         NUMBER OF SHOTS    =        2000
         NEW MAXIMUM WEIGHT =  1.1503371195500599E-03
         NEW MAXIMUM WEIGHT =  3.2720875047931022E-03
         NEW MAXIMUM WEIGHT =  3.4397725453424351E-02
         NEW MAXIMUM WEIGHT =  6.0381232770162795E-02
         NEW MAXIMUM WEIGHT =  6.6570674949068473E-02

         INITIAL SEARCH FINISHED

         OUTPUT ON ELEMENTARY PROCESS

         NUMBER OF EVENTS   =           0
         NUMBER OF WEIGHTS  =        2000
         MEAN VALUE OF WGT  =  4.5373E-03
         RMS SPREAD IN WGT  =  9.3312E-03
         ACTUAL MAX WEIGHT  =  6.0519E-02
         ASSUMED MAX WEIGHT =  6.6571E-02

         PROCESS CODE IPROC =       11706
         CROSS SECTION (PB) =   4.537    
         ERROR IN C-S  (PB) =  0.2087    
         EFFICIENCY PERCENT =   6.816

 EVENT     39:  900.00 GEV/C PBAR     ON  900.00 GEV/C P         PROCESS: 11706

 SEEDS:  875163092 &  655954870   STATUS: 100  ERROR:   0  WEIGHT: 0.4537E-02

                           ---INITIAL STATE---    

 IHEP    ID    IDPDG IST MO1 MO2 DA1 DA2    P-X     P-Y     P-Z   ENERGY   MASS 
   1 PBAR      -2212 101   0   0   0   0    0.00    0.00  900.00  900.00    0.94
   2 P          2212 102   0   0   0   0    0.00    0.00 -900.00  900.00    0.94
   3 CMF           0 103   1   2   0   0    0.00    0.00    0.00 1800.00 1800.00

                          ---HARD SUBPROCESS---   

 IHEP    ID    IDPDG IST MO1 MO2 DA1 DA2    P-X     P-Y     P-Z   ENERGY   MASS 
   4 UBAR         -2 121   6   7   9   5    0.00    0.00  312.09  312.09    0.32
   5 UQRK          2 122   6   4  17   8    0.00    0.00 -169.95  169.95    0.32
   6 HARD          0 120   4   5   7   8  -16.42   -3.93  142.14  482.34  460.61
   7 TBAR         -6 123   6   8  22   4  116.29  -61.69  157.43  266.49  170.00
   8 TQRK          6 124   6   5  24   7 -116.29   61.69  -15.29  215.55  170.00

                          ---PARTON SHOWERS---    

 IHEP    ID    IDPDG IST MO1 MO2 DA1 DA2    P-X     P-Y     P-Z   ENERGY   MASS 
   9 UBAR         94 141   4   6  11  16  -19.27   -6.00  314.16  310.83  -49.90
  10 CONE          0 100   4   7   0   0    0.88   -0.47    0.53    1.13    0.00
  11 UBARDBAR  -2101   2   9  12  45  21    0.00    0.00  408.95  408.95    0.70
  12 GLUON        21   2   9  13  46  47    8.42    0.19  140.64  140.89    0.75
  13 GLUON        21   2   9  14  48  49    2.07   -1.20   14.47   14.68    0.75
  14 DBAR         -1   2   9  15  50  49    3.78    3.25    8.85   10.16    0.32
  15 DQRK          1   2   9  16  51  50    3.65    2.24    9.47   10.40    0.32
  16 GLUON        21   2   9  26  52  53    1.36    1.52    3.46    4.09    0.75
  17 UQRK         94 142   5   6  19  21    2.85    2.07 -172.02  171.51  -13.73
  18 CONE          0 100   5   8   0   0   -0.88    0.47    0.07    1.00    0.00
  19 GLUON        21   2  17  20  54  55   -0.95   -0.97   -3.31    3.66    0.75
  20 GLUON        21   2  17  21  56  57   -1.90   -1.10  -16.01   16.17    0.75
  21 UD         2101   2  17  45  58  57    0.00    0.00 -708.66  708.66    1.04
  22 TBAR         94 143   7   6  23  23  107.70  -63.75  156.89  263.01  170.00
  23 TBAR         -6   3  22  22  26  26  107.70  -63.75  156.89  263.01  170.00
  24 TQRK         94 144   8   6  25  25 -124.12   59.82  -14.74  219.32  170.00
  25 TQRK          6   3  24  24  37  37 -124.12   59.82  -14.74  219.32  170.00

                       ---HEAVY FLAVOUR DECAYS--- 

 IHEP    ID    IDPDG IST MO1 MO2 DA1 DA2    P-X     P-Y     P-Z   ENERGY   MASS 
  26 TBAR         -6 155  22  37  27  29  107.70  -63.75  156.89  263.01  170.00
  27 MU-          13 123  26  28  30  28   18.31   32.76   65.37   75.38    0.11
  28 NU_MUBAR    -14 124  26  27  31  27   80.30  -57.83  106.04  145.04    0.00
  29 BBAR         -5 124  26  26  32  26    9.09  -38.68  -14.52   42.60    4.95
  30 MU-          13   1  27  26   0   0   17.82   31.88   63.62   73.36    0.11
  31 NU_MUBAR    -14   1  28  26   0   0   78.14  -56.28  103.19  141.14    0.00

                          ---PARTON SHOWERS---    

 IHEP    ID    IDPDG IST MO1 MO2 DA1 DA2    P-X     P-Y     P-Z   ENERGY   MASS 
  32 BBAR         94 144  29  26  34  36   11.74  -39.36   -9.92   48.52   23.85
  33 CONE          0 100  29  26   0   0    0.24    0.72    1.07    1.32    0.00
  34 GLUON        21   2  32  35  59  60   -2.95   -0.95   -3.35    4.62    0.75
  35 GLUON        21   2  32  36  61  62   -1.72   -1.41   -1.55    2.81    0.75
  36 BBAR         -5   2  32  44  63  62   16.41  -37.00   -5.02   41.08    4.95

                       ---HEAVY FLAVOUR DECAYS--- 

 IHEP    ID    IDPDG IST MO1 MO2 DA1 DA2    P-X     P-Y     P-Z   ENERGY   MASS 
  37 TQRK          6 155  24  19  38  40 -124.12   59.82  -14.74  219.32  170.00
  38 NU_E         12 123  37  39  41  39  -96.15   66.72   23.37  119.34    0.00
  39 E+          -11 124  37  38  42  38    6.38   13.33  -54.59   56.56    0.00
  40 BQRK          5 124  37  37  43  37  -34.36  -20.23   16.48   43.43    4.95
  41 NU_E         12   1  38  37   0   0  -96.15   66.72   23.37  119.34    0.00
  42 E+          -11   1  39  37   0   0    6.38   13.33  -54.59   56.56    0.00

                          ---PARTON SHOWERS---    

 IHEP    ID    IDPDG IST MO1 MO2 DA1 DA2    P-X     P-Y     P-Z   ENERGY   MASS 
  43 BQRK         94 144  40  37  44  44  -34.36  -20.23   16.48   43.43    4.95
  44 BQRK          5   2  43  54  64  63  -34.36  -20.23   16.48   43.43    4.95

                          ---GLUON SPLITTING---   

 IHEP    ID    IDPDG IST MO1 MO2 DA1 DA2    P-X     P-Y     P-Z   ENERGY   MASS 
  45 UBARDBAR  -2101 161   9  65  85  58    0.01    0.00  279.95  279.95    0.64
  46 UBAR         -2 158   9  47 104  84    1.90    0.01   33.44   33.50    0.32
  47 UQRK          2 158   9  69  86  46    3.96    0.11   64.98   65.11    0.32
  48 DBAR         -1 158   9  49  97  70    0.95   -0.68    7.08    7.18    0.32
  49 DQRK          1 158   9  71  87  48    0.40   -0.05    2.32    2.38    0.32
  50 DBAR         -1 158   9  51  98  72    3.00    2.57    7.04    8.08    0.32
  51 DQRK          1 158   9  52  88  50    3.65    2.24    9.47   10.40    0.32
  52 DBAR         -1 158   9  53  88  51    0.49    0.47    0.95    1.21    0.32
  53 DQRK          1 158   9  73  89  52    0.79    0.96    2.29    2.62    0.32
  54 DBAR         -1 158  17  55 102  80   -0.23   -0.13   -0.54    0.68    0.32
  55 DQRK          1 158  17  56  90  54   -0.62   -0.80   -2.35    2.58    0.32
  56 DBAR         -1 158  17  57  90  55   -1.18   -0.54   -8.28    8.38    0.32
  57 DQRK          1 158  17  75  91  56   -0.34   -0.26   -5.03    5.06    0.32
  58 UD         2101 162  17  45  96  68    0.00    0.00 -552.77  552.77    0.64
  59 DBAR         -1 158  32  60  99  74   -0.85   -0.40   -0.95    1.37    0.32
  60 DQRK          1 158  32  61  92  59   -1.65   -0.34   -1.90    2.56    0.32
  61 DBAR         -1 158  32  62  92  60   -0.66   -0.74   -0.83    1.33    0.32
  62 DQRK          1 158  32  77  93  61   -0.91   -0.59   -0.62    1.29    0.32
  63 BBAR         -5 158  32  64 101  78   14.03  -31.87   -4.37   35.44    4.95
  64 BQRK          5 158  43  81  94  63  -24.17  -14.23   11.39   30.68    4.95
  65 DBAR         -1 159   9  66  85  45    0.06    0.00   30.61   30.61    0.32
  66 DQRK          1 159   9  83  95  65    0.02    0.00   65.93   65.93    0.32
  67 UBAR         -2 159  17  68 100  76    0.00    0.00 -108.31  108.31    0.32
  68 UQRK          2 159  17  58  96  67   -0.02   -0.02  -26.64   26.65    0.32
  69 DBAR         -1 159   9  70  86  47    0.44   -0.20    4.71    4.75    0.32
  70 DQRK          1 159   9  48  97  69    2.01    0.04   32.91   32.97    0.32
  71 SBAR         -3 159   9  72  87  49    0.67    0.43    2.09    2.29    0.50
  72 SQRK          3 159   9  50  98  71    0.55    0.00    2.95    3.04    0.50
  73 UBAR         -2 159  32  74  89  53   -0.35   -0.15   -0.36    0.61    0.32
  74 UQRK          2 159   9  59  99  73   -0.02    0.04    0.09    0.33    0.32
  75 SBAR         -3 159  17  76  91  57   -0.10   -0.08  -15.37   15.37    0.50
  76 SQRK          3 159  17  67 100  75   -0.26   -0.20   -8.28    8.30    0.50
  77 SBAR         -3 159  32  78  93  62    1.73   -3.90   -0.53    4.33    0.50
  78 SQRK          3 159  32  63 101  77    0.49   -1.31   -0.22    1.50    0.50
  79 DBAR         -1 159  17  80 103  82   -0.22   -0.12   -0.18    0.45    0.32
  80 DQRK          1 159  43  54 102  79   -3.20   -1.89    1.55    4.04    0.32
  81 UBAR         -2 159  17  82  94  64   -1.26   -0.74    0.58    1.61    0.32
  82 UQRK          2 159  43  79 103  81   -5.60   -3.30    2.72    7.05    0.32
  83 DBAR         -1 159   9  84  95  66    0.07    0.00   27.33   27.33    0.32
  84 DQRK          1 159   9  46 104  83    0.23    0.00   11.57   11.58    0.32

                         ---CLUSTER FORMATION---  

 IHEP    ID    IDPDG IST MO1 MO2 DA1 DA2    P-X     P-Y     P-Z   ENERGY   MASS 
  85 CLUS         91 184  45  65 105 106    0.07    0.00  310.56  310.56    1.23
  86 CLUS         91 183  47  69 107 108    4.40   -0.09   69.69   69.85    1.59
  87 CLUS         91 183  49  71 109 110    1.07    0.38    4.41    4.67    1.03
  88 CLUS         91 183  51  52 111 112    4.14    2.70   10.42   11.61    1.31
  89 CLUS         91 183  53  73 113 114    0.44    0.80    1.92    3.24    2.44
  90 CLUS         91 183  55  56 115 116   -1.80   -1.34  -10.63   10.96    1.48
  91 CLUS         91 183  57  75 117 118   -0.44   -0.34  -20.39   20.43    1.09
  92 CLUS         91 183  60  61 119 120   -2.31   -1.08   -2.73    3.89    1.09
  93 CLUS         91 183  62  77 121 122    0.82   -4.49   -1.15    5.62    3.07
  94 CLUS         91 183  64  81 123 123  -25.44  -14.98   11.97   32.29    5.28
  95 CLUS         91 183  66  83 124 125    0.09    0.00   93.26   93.26    0.71
  96 CLUS         91 185  68  58 126 127   -0.02   -0.02 -579.41  579.41    1.64
  97 CLUS         91 183  70  48 128 129    2.97   -0.64   39.98   40.15    2.04
  98 CLUS         91 183  72  50 130 131    3.54    2.57    9.99   11.12    2.17
  99 CLUS         91 183  74  59 132 133   -0.87   -0.36   -0.86    1.71    1.13
 100 CLUS         91 183  76  67 134 135   -0.26   -0.20 -116.59  116.61    2.24
 101 CLUS         91 183  78  63 136 136   14.56  -33.17   -4.60   36.91    5.38
 102 CLUS         91 183  80  54 137 138   -3.47   -2.02    1.02    4.76    2.34
 103 CLUS         91 183  82  79 139 140   -5.81   -3.42    2.54    7.49    2.06
 104 CLUS         91 183  84  46 141 142    2.13    0.01   45.01   45.08    1.04

                          ---CLUSTER DECAYS---    

 IHEP    ID    IDPDG IST MO1 MO2 DA1 DA2    P-X     P-Y     P-Z   ENERGY   MASS 
 105 PBAR      -2212   1  85   9   0   0   -0.13    0.09  215.09  215.09    0.94
 106 PI+         211   1  85   9   0   0    0.20   -0.09   95.47   95.47    0.14
 107 OMEGA       223 197  86   9 143 145    2.33   -0.03   34.12   34.20    0.78
 108 RHO+        213 197  86   9 146 147    2.07   -0.06   35.58   35.65    0.77
 109 PI0         111   1  87   9   0   0    0.14    0.05    0.57    0.60    0.14
 110 K*0         313 197  87   9 148 149    0.93    0.34    3.84    4.07    0.90
 111 PI0         111   1  88   9   0   0    2.48    1.51    6.44    7.07    0.14
 112 OMEGA       223 197  88   9 150 151    1.66    1.19    3.98    4.54    0.78
 113 P          2212   1  89   9   0   0   -0.35    0.36    0.89    1.39    0.94
 114 DLTABR--  -2224 197  89   9 152 153    0.80    0.45    1.03    1.85    1.23
 115 A_10      20113 197  90  17 154 155   -1.73   -1.30  -10.33   10.63    1.23
 116 PI0         111   1  90  17   0   0   -0.06   -0.04   -0.30    0.33    0.14
 117 PI-        -211   1  91  17   0   0   -0.08    0.11  -12.23   12.23    0.14
 118 K+          321   1  91  17   0   0   -0.36   -0.44   -8.16    8.20    0.49
 119 RHO-       -213 197  92  32 156 157   -1.94   -1.07   -2.51    3.44    0.77
 120 PI+         211   1  92  32   0   0   -0.37    0.00   -0.22    0.45    0.14
 121 KL_10     10313 197  93  32 158 159    1.17   -2.31   -1.07    3.22    1.57
 122 ETAP        331 197  93  32 160 162   -0.35   -2.18   -0.08    2.41    0.96
 123 B-         -521 196  94  43 163 165  -25.44  -14.98   11.97   32.29    5.28
 124 PI0         111   1  95   9   0   0    0.30   -0.06   25.24   25.24    0.14
 125 PI0         111   1  95   9   0   0   -0.21    0.06   68.02   68.02    0.14
 126 PI+         211   1  96  17   0   0    0.04    0.14 -231.52  231.52    0.14
 127 DELTA0     2114 197  96  17 166 167   -0.06   -0.15 -347.89  347.89    1.23
 128 P          2212   1  97   9   0   0    0.84   -0.11   14.92   14.97    0.94
 129 PBAR      -2212   1  97   9   0   0    2.13   -0.53   25.06   25.17    0.94
 130 ETA         221 197  98   9 168 170    0.59    0.27    2.25    2.40    0.55
 131 K*_2BAR0   -315 197  98   9 171 172    2.95    2.30    7.74    8.72    1.43
 132 PI0         111   1  99   9   0   0   -0.35    0.05   -0.95    1.02    0.14
 133 PI+         211   1  99   9   0   0   -0.52   -0.41    0.09    0.68    0.14
 134 KBAR0      -311 197 100  17 173 173    0.04   -0.01  -17.14   17.15    0.50
 135 PI_2-    -10215 197 100  17 174 175   -0.30   -0.19  -99.44   99.46    1.67
 136 B_S0        531 200 101  32 176 176   14.56  -33.17   -4.60   36.91    5.38
 137 HL_10     10223 197 102  43 177 178   -3.20   -1.92    1.24    4.10    1.17
 138 ETA         221 197 102  43 179 181   -0.28   -0.10   -0.22    0.66    0.55
 139 A_20        115 197 103  43 182 184   -2.96   -1.35    1.04    3.66    1.32
 140 PI+         211   1 103  43   0   0   -2.85   -2.07    1.50    3.84    0.14
 141 PI0         111   1 104   9   0   0    0.66    0.06   16.64   16.65    0.14
 142 RHO-       -213 197 104   9 185 186    1.47   -0.05   28.37   28.42    0.77

                       ---STRONG HADRON DECAYS--- 

 IHEP    ID    IDPDG IST MO1 MO2 DA1 DA2    P-X     P-Y     P-Z   ENERGY   MASS 
 143 PI+         211   1 107   9   0   0    1.90   -0.02   24.27   24.35    0.14
 144 PI-        -211   1 107   9   0   0    0.32    0.00    6.77    6.78    0.14
 145 PI0         111   1 107   9   0   0    0.11   -0.02    3.07    3.08    0.14
 146 PI+         211   1 108   9   0   0    1.80    0.19   29.01   29.07    0.14
 147 PI0         111   1 108   9   0   0    0.27   -0.25    6.57    6.58    0.14
 148 K0          311 198 110   9 187 187    0.65    0.06    3.19    3.29    0.50
 149 PI0         111   1 110   9   0   0    0.28    0.27    0.65    0.77    0.14
 150 PI0         111   1 112   9   0   0    1.33    1.12    3.73    4.12    0.14
 151 GAMMA        22   1 112   9   0   0    0.34    0.07    0.25    0.42    0.00
 152 PBAR      -2212   1 114   9   0   0    0.75    0.24    1.02    1.59    0.94
 153 PI-        -211   1 114   9   0   0    0.05    0.21    0.01    0.25    0.14
 154 RHO+        213 198 115  17 188 189   -1.57   -0.83   -8.84    9.05    0.77
 155 PI-        -211   1 115  17   0   0   -0.16   -0.47   -1.49    1.58    0.14
 156 PI-        -211   1 119  32   0   0   -0.65   -0.64   -1.38    1.66    0.14
 157 PI0         111   1 119  32   0   0   -1.29   -0.43   -1.12    1.77    0.14
 158 K+          321   1 121  32   0   0    0.56   -0.88    0.04    1.15    0.49
 159 RHO-       -213 198 121  32 190 191    0.61   -1.43   -1.11    2.06    0.77
 160 PI+         211   1 122  32   0   0   -0.15   -0.86    0.05    0.89    0.14
 161 PI-        -211   1 122  32   0   0   -0.07   -0.48   -0.08    0.51    0.14
 162 ETA         221 198 122  32 192 193   -0.13   -0.84   -0.05    1.01    0.55
 163 RHO0        113 198 123  43 194 195  -18.44   -9.34    6.79   21.77    0.77
 164 E-           11   1 123  43   0   0   -6.49   -5.26    5.27    9.88    0.00
 165 NU_EBAR     -12   1 123  43   0   0   -0.50   -0.38   -0.09    0.63    0.00
 166 P          2212   1 127  17   0   0    0.09   -0.11 -219.33  219.33    0.94
 167 PI-        -211   1 127  17   0   0   -0.15   -0.05 -128.56  128.56    0.14
 168 PI0         111   1 130   9   0   0    0.07    0.07    0.80    0.82    0.14
 169 PI0         111   1 130   9   0   0    0.21    0.13    0.47    0.55    0.14
 170 PI0         111   1 130   9   0   0    0.31    0.07    0.98    1.04    0.14
 171 K-         -321   1 131   9   0   0    2.36    1.08    5.22    5.85    0.49
 172 PI+         211   1 131   9   0   0    0.59    1.22    2.53    2.87    0.14
 173 K_S0        310 198 134  17 196 197    0.04   -0.01  -17.14   17.15    0.50
 174 F_2         225 198 135  17 198 199   -0.33   -0.32  -95.89   95.90    1.27
 175 PI-        -211   1 135  17   0   0    0.03    0.12   -3.56    3.56    0.14
 176 B_SBAR0    -531 199 136  32 207 208   14.56  -33.17   -4.60   36.91    5.38
 177 RHO+        213 198 137  43 200 201   -2.29   -1.73    0.99    3.13    0.77
 178 PI-        -211   1 137  43   0   0   -0.90   -0.19    0.24    0.96    0.14
 179 PI0         111   1 138  43   0   0    0.01    0.05    0.02    0.15    0.14
 180 PI0         111   1 138  43   0   0   -0.07   -0.10   -0.06    0.19    0.14
 181 PI0         111   1 138  43   0   0   -0.22   -0.05   -0.18    0.32    0.14
 182 OMEGA       223 198 139  43 202 204   -1.92   -0.63    0.73    2.28    0.78
 183 PI+         211   1 139  43   0   0   -0.66   -0.32    0.05    0.75    0.14
 184 PI-        -211   1 139  43   0   0   -0.38   -0.40    0.26    0.63    0.14
 185 PI-        -211   1 142   9   0   0    0.68   -0.38   13.27   13.29    0.14
 186 PI0         111   1 142   9   0   0    0.79    0.33   15.11   15.13    0.14
 187 K_S0        310 198 148   9 205 206    0.65    0.06    3.19    3.29    0.50
 188 PI+         211   1 154  17   0   0   -1.45   -0.73   -8.47    8.62    0.14
 189 PI0         111   1 154  17   0   0   -0.13   -0.10   -0.37    0.43    0.14
 190 PI-        -211   1 159  32   0   0    0.68   -1.24   -1.13    1.82    0.14
 191 PI0         111   1 159  32   0   0   -0.06   -0.19    0.02    0.24    0.14
 192 GAMMA        22   1 162  32   0   0   -0.31   -0.62   -0.15    0.71    0.00
 193 GAMMA        22   1 162  32   0   0    0.18   -0.22    0.10    0.30    0.00
 194 PI+         211   1 163  43   0   0  -16.47   -8.12    6.03   19.32    0.14
 195 PI-        -211   1 163  43   0   0   -1.98   -1.22    0.76    2.45    0.14
 196 PI0         111   1 173  17   0   0    0.14    0.16   -6.36    6.37    0.14
 197 PI0         111   1 173  17   0   0   -0.10   -0.16  -10.78   10.78    0.14
 198 PI0         111   1 174  17   0   0   -0.61   -0.52  -38.85   38.86    0.14
 199 PI0         111   1 174  17   0   0    0.27    0.20  -57.03   57.04    0.14
 200 PI+         211   1 177  43   0   0   -1.01   -0.34    0.30    1.11    0.14
 201 PI0         111   1 177  43   0   0   -1.29   -1.39    0.70    2.02    0.14
 202 PI+         211   1 182  43   0   0   -0.30   -0.02    0.08    0.34    0.14
 203 PI-        -211   1 182  43   0   0   -0.30   -0.13    0.36    0.50    0.14
 204 PI0         111   1 182  43   0   0   -1.32   -0.48    0.29    1.44    0.14
 205 PI0         111   1 187   9   0   0    0.51   -0.06    1.56    1.65    0.14
 206 PI0         111   1 187   9   0   0    0.14    0.12    1.63    1.64    0.14

                       ---HEAVY FLAVOUR DECAYS--- 

 IHEP    ID    IDPDG IST MO1 MO2 DA1 DA2    P-X     P-Y     P-Z   ENERGY   MASS 
 207 BQRK          5 155 176 208 209 211   13.20  -30.08   -4.17   33.47    4.88
 208 SBAR         -3 125 176 211 212 211    1.35   -3.09   -0.43    3.43    0.50
 209 CQRK          4 123 207 210 213 210    2.30   -5.94   -0.61    6.59    1.55
 210 CBAR         -4 124 207 209 215 209    3.58   -8.37   -1.74    9.40    1.55
 211 SQRK          3 124 207 207 217 207    7.33  -15.77   -1.82   17.49    0.50
 212 SBAR         -3 160 208 221 223 221    1.35   -3.09   -0.43    3.43    0.50

                          ---PARTON SHOWERS---    

 IHEP    ID    IDPDG IST MO1 MO2 DA1 DA2    P-X     P-Y     P-Z   ENERGY   MASS 
 213 CQRK         94 143 209 207 214 214    2.30   -5.94   -0.61    6.59    1.55
 214 CQRK          4   2 213 216 219 216    2.30   -5.94   -0.61    6.59    1.55
 215 CBAR         94 144 210 207 216 216    3.58   -8.37   -1.74    9.40    1.55
 216 CBAR         -4   2 215 219 220 219    3.58   -8.37   -1.74    9.40    1.55
 217 SQRK         94 144 211 207 218 218    7.33  -15.77   -1.82   17.49    0.50
 218 SQRK          3   2 217 212 221 212    7.33  -15.77   -1.82   17.49    0.50

                          ---GLUON SPLITTING---   

 IHEP    ID    IDPDG IST MO1 MO2 DA1 DA2    P-X     P-Y     P-Z   ENERGY   MASS 
 219 CQRK          4 158 213 220 222 220    2.30   -5.94   -0.61    6.59    1.55
 220 CBAR         -4 158 215 219 222 219    3.58   -8.37   -1.74    9.40    1.55
 221 SQRK          3 158 217 212 223 212    7.33  -15.77   -1.82   17.49    0.50

                         ---CLUSTER FORMATION---  

 IHEP    ID    IDPDG IST MO1 MO2 DA1 DA2    P-X     P-Y     P-Z   ENERGY   MASS 
 222 CLUS         91 183 219 220 224 224    4.87  -12.18   -2.12   13.62    2.98
 223 CLUS         91 183 221 212 225 226    9.69  -20.99   -2.48   23.29    1.37

                          ---CLUSTER DECAYS---    

 IHEP    ID    IDPDG IST MO1 MO2 DA1 DA2    P-X     P-Y     P-Z   ENERGY   MASS 
 224 ETA_C       441 199 222 213 227 229    4.87  -12.18   -2.12   13.62    2.98
 225 K-         -321   1 223 217   0   0    8.22  -17.67   -2.03   19.60    0.49
 226 K+          321   1 223 217   0   0    1.47   -3.31   -0.45    3.69    0.49

                       ---HEAVY FLAVOUR DECAYS--- 

 IHEP    ID    IDPDG IST MO1 MO2 DA1 DA2    P-X     P-Y     P-Z   ENERGY   MASS 
 227 GLUON        21 123 224 229 230 228    2.59   -4.45   -1.11    5.32    0.75
 228 GLUON        21 124 224 227 232 229    0.84   -4.20   -0.51    4.38    0.75
 229 GLUON        21 124 224 228 234 227    1.44   -3.53   -0.50    3.92    0.75

                          ---PARTON SHOWERS---    

 IHEP    ID    IDPDG IST MO1 MO2 DA1 DA2    P-X     P-Y     P-Z   ENERGY   MASS 
 230 GLUON        94 143 227 224 231 231    2.59   -4.45   -1.11    5.32    0.75
 231 GLUON        21   2 230 235 236 237    2.59   -4.45   -1.11    5.32    0.75
 232 GLUON        94 144 228 224 233 233    0.84   -4.20   -0.51    4.38    0.75
 233 GLUON        21   2 232 236 238 239    0.84   -4.20   -0.51    4.38    0.75
 234 GLUON        94 144 229 224 235 235    1.44   -3.53   -0.50    3.92    0.75
 235 GLUON        21   2 234 238 240 241    1.44   -3.53   -0.50    3.92    0.75

                          ---GLUON SPLITTING---   

 IHEP    ID    IDPDG IST MO1 MO2 DA1 DA2    P-X     P-Y     P-Z   ENERGY   MASS 
 236 DBAR         -1 158 230 237 243 239    1.65   -2.62   -0.82    3.22    0.32
 237 DQRK          1 158 230 240 242 236    0.95   -1.83   -0.29    2.10    0.32
 238 DBAR         -1 158 232 239 244 241    0.66   -3.17   -0.34    3.27    0.32
 239 DQRK          1 158 232 236 243 238    0.18   -1.03   -0.17    1.11    0.32
 240 UBAR         -2 158 234 241 242 237    0.65   -2.08   -0.24    2.22    0.32
 241 UQRK          2 158 234 238 244 240    0.79   -1.45   -0.26    1.70    0.32

                         ---CLUSTER FORMATION---  

 IHEP    ID    IDPDG IST MO1 MO2 DA1 DA2    P-X     P-Y     P-Z   ENERGY   MASS 
 242 CLUS         91 183 237 240 245 246    1.60   -3.90   -0.53    4.32    0.74
 243 CLUS         91 183 239 236 247 248    1.82   -3.65   -0.98    4.33    1.03
 244 CLUS         91 183 241 238 249 250    1.45   -4.62   -0.60    4.98    0.96

                          ---CLUSTER DECAYS---    

 IHEP    ID    IDPDG IST MO1 MO2 DA1 DA2    P-X     P-Y     P-Z   ENERGY   MASS 
 245 PI-        -211   1 242 230   0   0    1.53   -3.79   -0.46    4.11    0.14
 246 PI0         111   1 242 230   0   0    0.06   -0.12   -0.07    0.20    0.14
 247 PI0         111   1 243 232   0   0    0.52   -0.39    0.03    0.66    0.14
 248 PI0         111   1 243 232   0   0    1.30   -3.27   -1.01    3.66    0.14
 249 PI0         111   1 244 234   0   0    1.02   -3.98   -0.74    4.18    0.14
 250 PI+         211   1 244 234   0   0    0.43   -0.64    0.14    0.80    0.14

         OUTPUT ON ELEMENTARY PROCESS

         NUMBER OF EVENTS   =        1000
         NUMBER OF WEIGHTS  =       14518
         MEAN VALUE OF WGT  =  4.4894E-03
         RMS SPREAD IN WGT  =  9.2221E-03
         ACTUAL MAX WEIGHT  =  6.1048E-02
         ASSUMED MAX WEIGHT =  6.6571E-02

         PROCESS CODE IPROC =       11706
         CROSS SECTION (PB) =   4.489    
         ERROR IN C-S  (PB) =  7.6538E-02
         EFFICIENCY PERCENT =   6.744    

-----------------------------------------------------------------------

               ****** 22. GUIDE TO SAMPLE OUTPUT ******

   After listing  the more important input parameter values, the program
   prints the message

         NO EVENTS WILL BE WRITTEN TO DISK

   to remind the user that LWEVT=0 for this run. Since BBbar oscillation
   is enabled (MIXING=.TRUE.), the relevant parameters are printed:

         B_d: Delt-M/Gam =0.7000 Delt-Gam/2*Gam =0.0000
         B_s: Delt-M/Gam = 10.00 Delt-Gam/2*Gam =0.2000

   The messages

         PDFLIB NOT USED FOR BEAM 1
         PDFLIB NOT USED FOR BEAM 2

   indicating that the  CERN PDFLIB  structure function library will not
   be used  (MODPDF<0).  Next the particle property and decay tables are
   checked for consistency.  The messages

Line,  565 decay: LMBDA_C+ --> XI*0     K*+                                
is kinematically not allowed, Min-Mout=     -0.139
LMBDA_C+: BR sum = 0.97800
Rescaling to 1

Line,  990 decay: LMBDA_C- --> XI*BAR   K*-                                
is kinematically not allowed, Min-Mout=     -0.139
LMBDA_C-: BR sum = 0.97800
Rescaling to 1

   indicate that some user-modified decay modes are impossible  and will
   be ignored.  The default particle data table  was modified by calling
   HWUSTA('PI0     ') to suppress pi0 decays, so we get the message

        PARTICLE TYPE  21=PI0  SET STABLE

   Next the program  searches for the  maximum weight,  i.e. the maximum
   cross section in the available phase space, as implied by the default
   value WGTMAX=0. The parameters

        MIN P-TRAN FOR 2->2    =   10.0000
        MAX P-TRAN FOR 2->2    =  900.0002
   with
        PROCESS CODE       =       11706

   mean that the transverse momentum of the t quark in the QCD 2->2 hard
   subprocesses  is required  to be  greater than  10 GeV/c.  After this
   search, the estimated total cross section of relevant subprocesses in
   this region of phase space is printed,  together with the anticipated
   efficiency of subprocess generation (i.e. average/maximum weight):

         CROSS SECTION (PB) =   4.537    
         ERROR IN C-S  (PB) =  0.2087    
         EFFICIENCY PERCENT =   6.816    

   Since the  print  parameter  was  MAXPR=0,  no events were printed by 
   default,  but the user analysis routine HWANAL called HWUEPR to print
   the first "interesting" event. The event heading

EVENT     39:  900.00 GEV/C PBAR     ON  900.00 GEV/C P         PROCESS: 11706

SEEDS:  875163092 &  655954870   STATUS: 100  ERROR:   0  WEIGHT: 0.4537E-02

   tells us  the beam and target, the random number seeds at the start of
   the  event and the process code IPROC. The status 100 means a complete
   event  was generated  and the zero  error code means  no problems were
   encountered.  Since  NOWGT=.TRUE. (unweighted event generation),  each
   event has the mean weight computed earlier.

   Next come the contents of  COMMON/HEPEVT/ and related quantities. The
   print parameter for vertex information has been set PRVTX=.FALSE. and
   so no  space-time information  is printed.  The various parts of this
   particular event are located as follows:

             +---------+--------------------------------------+
             |   Entry | Description                          |
             +---------+--------------------------------------+
             |   1-  3 | Initial state                        |
             |   4-  8 | Hard subprocess: u+ubar -> t+tbar    |
             |   9- 25 | Parton showers                       |
             |  26- 44 | Top decays and subsequent showers    |
             |  45- 84 | Gluon splitting                      |
             |  85-104 | Cluster formation                    |
             | 105-206 | Cluster and hadron decays            |
             | 207-218 | Weak decay of B_sbar and showers     |
             | 219-226 | Hadronization of B_sbar products     |
             | 227-235 | 3-gluon decay of eta_c               |
             | 236-250 | Hadronization of eta_c products      |
             +---------+--------------------------------------+

   We discuss each part in turn.

                           ---INITIAL STATE---    

 IHEP    ID    IDPDG IST MO1 MO2 DA1 DA2    P-X     P-Y     P-Z   ENERGY   MASS 
   1 PBAR      -2212 101   0   0   0   0    0.00    0.00  900.00  900.00    0.94
   2 P          2212 102   0   0   0   0    0.00    0.00 -900.00  900.00    0.94
   3 CMF           0 103   1   2   0   0    0.00    0.00    0.00 1800.00 1800.00

   CMF represents  the overall centre of mass of the initial state.  The
   'mother' MOi=JMOHEP(i,IHEP) & 'daughter' DAi=JDAHEP(i,IHEP)  pointers
   are set to zero for these entries.

                          ---HARD SUBPROCESS---   

 IHEP    ID    IDPDG IST MO1 MO2 DA1 DA2    P-X     P-Y     P-Z   ENERGY   MASS 
   4 UBAR         -2 121   6   7   9   5    0.00    0.00  312.09  312.09    0.32
   5 UQRK          2 122   6   4  17   8    0.00    0.00 -169.95  169.95    0.32
   6 HARD          0 120   4   5   7   8  -16.42   -3.93  142.14  482.34  460.61
   7 TBAR         -6 123   6   8  22   4  116.29  -61.69  157.43  266.49  170.00
   8 TQRK          6 124   6   5  24   7 -116.29   61.69  -15.29  215.55  170.00

   HARD is the hard subprocess  centre of mass.  Its mother and daughter
   pointers give the locations of the incoming and outgoing partons. The
   status codes 121-124 correspond to the  hard subprocess  partons 1-4.
   The first mother pointers show the location of the hard c.m., and the
   second mother of each parton  is the  'colour  mother',  as explained
   above. Thus the colours of partons 1234 are connected to 3142 respt.,
   corresponding to process IHPRO=12. Likewise,the first daughter points 
   to the associated jet but the second daughter is the colour daughter,
   i.e. the parton to which this one's anticolour is connected. Thus the
   anticolour connections of 1234 in this case are to 2413.   The colour
   diagram is

                     (ubar)1 --<--+     +--<-- 3(tbar)
                                   \___/
                                    ___
                                   /   \
                        (u)2 -->--+     +-->-- 4(t)

   Note that in specifying the colour connections all lines are regarded
   as outgoing, and that since antiquarks carry no colour MO2 is in that
   case used for the flavour connection (similarly with DA2 for quarks).
   Gluon radiation from the initial ubar will be limited by interference
   with the tbar and vice-versa,  that from the incoming  u by the t and
   vice-versa.  At this stage, the momenta and masses of the partons are
   the raw on-shell values  generated before QCD radiative  corrections,
   but HARD has been updated  to give the true hard  subprocess momentum
   after initial- and final-state parton branching.

                          ---PARTON SHOWERS---    

   The QCD cascade from each hard parton is generated in sequence. First
   there is a jet entry (IDHEP=94)  giving the total jet momentum,  mass
   and flavour. For initial-state jets the mass represents  -|q**2|**1/2
   for the virtual parton  entering the  hard subprocess.  MO1 gives the
   parent hard parton and MO2 the hard centre-of-mass. DO1 and DO2 point
   to the first and last parton in the jet after perturbative branching.
   If branching occurs, the next entry  (CONE)  is a lightlike  4-vector
   defining the  radiation  cone  and the  orientation of  the radiation
   pattern.

   The partons in the jet (with ISTHEP set to 2 by  gluon splitting sub-
   routine  HWCGSP)  have their colour and anticolour  connections given
   by MO2 and DA2  respectively,  as described for  the hard subprocess.
   For an incoming jet, the remnants of the incoming hadrons (IHEP=11,21
   here) also have ISTHEP=2.  The ubar jet is:

 IHEP    ID    IDPDG IST MO1 MO2 DA1 DA2    P-X     P-Y     P-Z   ENERGY   MASS 
   9 UBAR         94 141   4   6  11  16  -19.27   -6.00  314.16  310.83  -49.90
  10 CONE          0 100   4   7   0   0    0.88   -0.47    0.53    1.13    0.00
  11 UBARDBAR  -2101   2   9  12  45  21    0.00    0.00  408.95  408.95    0.70
  12 GLUON        21   2   9  13  46  47    8.42    0.19  140.64  140.89    0.75
  13 GLUON        21   2   9  14  48  49    2.07   -1.20   14.47   14.68    0.75
  14 DBAR         -1   2   9  15  50  49    3.78    3.25    8.85   10.16    0.32
  15 DQRK          1   2   9  16  51  50    3.65    2.24    9.47   10.40    0.32
  16 GLUON        21   2   9  26  52  53    1.36    1.52    3.46    4.09    0.75

   and similarly for the u jet (IHEP=17-21). The produced t and tbar are
   so slow in the subprocess c.o.m. frame  that they do not radiate  any
   resolvable gluons. After any showering, they're given status ISTHEP=3
   and copied with ISTHEP=155 retaining the colour connection labels for
   the decay processes. In this event both top decays are leptonic:

                        ---HEAVY FLAVOUR DECAYS--- 

 IHEP    ID    IDPDG IST MO1 MO2 DA1 DA2    P-X     P-Y     P-Z   ENERGY   MASS 
  26 TBAR         -6 155  22  37  27  29  107.70  -63.75  156.89  263.01  170.00
  27 MU-          13 123  26  28  30  28   18.31   32.76   65.37   75.38    0.11
  28 NU_MUBAR    -14 124  26  27  31  27   80.30  -57.83  106.04  145.04    0.00
  29 BBAR         -5 124  26  26  32  26    9.09  -38.68  -14.52   42.60    4.95
  30 MU-          13   1  27  26   0   0   17.82   31.88   63.62   73.36    0.11
  31 NU_MUBAR    -14   1  28  26   0   0   78.14  -56.28  103.19  141.14    0.00

  37 TQRK          6 155  24  19  38  40 -124.12   59.82  -14.74  219.32  170.00
  38 NU_E         12 123  37  39  41  39  -96.15   66.72   23.37  119.34    0.00
  39 E+          -11 124  37  38  42  38    6.38   13.33  -54.59   56.56    0.00
  40 BQRK          5 124  37  37  43  37  -34.36  -20.23   16.48   43.43    4.95
  41 NU_E         12   1  38  37   0   0  -96.15   66.72   23.37  119.34    0.00
  42 E+          -11   1  39  37   0   0    6.38   13.33  -54.59   56.56    0.00

                        ---PARTON SHOWERS---    

   After the tbar decay, the resulting bbar radiates 2 gluons:

 IHEP    ID    IDPDG IST MO1 MO2 DA1 DA2    P-X     P-Y     P-Z   ENERGY   MASS 
  32 BBAR         94 144  29  26  34  36   11.74  -39.36   -9.92   48.52   23.85
  33 CONE          0 100  29  26   0   0    0.24    0.72    1.07    1.32    0.00
  34 GLUON        21   2  32  35  59  60   -2.95   -0.95   -3.35    4.62    0.75
  35 GLUON        21   2  32  36  61  62   -1.72   -1.41   -1.55    2.81    0.75
  36 BBAR         -5   2  32  44  63  62   16.41  -37.00   -5.02   41.08    4.95

   but the b quark from the t decay does not radiate.  If the decays had
   been hadronic, the quarks from the virtual W decay would also radiate
   in general.

                          ---GLUON SPLITTING---   

   As the first step  in the cluster hadronization model,  any gluons in
   the jets are split into light quark-antiquark pairs.  The flavours of
   the pairs are  chosen at random  amongst those allowed by kinematics.
   The colour connections are remade accordingly.

 IHEP    ID    IDPDG IST MO1 MO2 DA1 DA2    P-X     P-Y     P-Z   ENERGY   MASS 
  45 UBARDBAR  -2101 161   9  65  85  58    0.01    0.00  279.95  279.95    0.64
  46 UBAR         -2 158   9  47 104  84    1.90    0.01   33.44   33.50    0.32
  47 UQRK          2 158   9  69  86  46    3.96    0.11   64.98   65.11    0.32
  48 DBAR         -1 158   9  49  97  70    0.95   -0.68    7.08    7.18    0.32
  49 DQRK          1 158   9  71  87  48    0.40   -0.05    2.32    2.38    0.32
.......
  63 BBAR         -5 158  32  64 101  78   14.03  -31.87   -4.37   35.44    4.95
  64 BQRK          5 158  43  81  94  63  -24.17  -14.23   11.39   30.68    4.95

   Each quark  (or antidiquark) is combined with its colour mother anti-
   quark (or diquark) to make a cluster with the sum of their 4-momenta.
   All  non-beam  clusters with  masses above  the maximum  are split by
   creating new quark-antiquark pairs with  ISTHEP=159 (10 such pairs in
   this event). 

  65 DBAR         -1 159   9  66  85  45    0.06    0.00   30.61   30.61    0.32
  66 DQRK          1 159   9  83  95  65    0.02    0.00   65.93   65.93    0.32
.......
  83 DBAR         -1 159   9  84  95  66    0.07    0.00   27.33   27.33    0.32
  84 DQRK          1 159   9  46 104  83    0.23    0.00   11.57   11.58    0.32

                         ---CLUSTER FORMATION---  

   Next the clusters themselves are listed. The mothers of a cluster are
   the partons from which it is made,  and the daughters are the primary
   hadrons into which it decays.

 IHEP    ID    IDPDG IST MO1 MO2 DA1 DA2    P-X     P-Y     P-Z   ENERGY   MASS 
  85 CLUS         91 184  45  65 105 106    0.07    0.00  310.56  310.56    1.23
  86 CLUS         91 183  47  69 107 108    4.40   -0.09   69.69   69.85    1.59
.......
 103 CLUS         91 183  82  79 139 140   -5.81   -3.42    2.54    7.49    2.06
 104 CLUS         91 183  84  46 141 142    2.13    0.01   45.01   45.08    1.04

                          ---CLUSTER DECAYS---    

   The  clusters,  including the b-flavoured  clusters  94 and 101,  now
   decay,  usually into pairs of hadrons chosen according to the density
   of states.  Sometimes  single-hadron  decays occur,  with transfer of
   momentum to  a neighbouring cluster,  if there is  insufficient phase
   space for two-body decay.  Note that cluster 94 actually did a 1-body
   decay into a B- (IHEP=123,  ISTHEP=196).  Hadrons with  ISTHEP=1  are
   stable.  ISTHEP=200  indicates a  neutral  B meson  which may undergo
   flavour oscillation.

 IHEP    ID    IDPDG IST MO1 MO2 DA1 DA2    P-X     P-Y     P-Z   ENERGY   MASS 
 105 PBAR      -2212   1  85   9   0   0   -0.13    0.09  215.09  215.09    0.94
 106 PI+         211   1  85   9   0   0    0.20   -0.09   95.47   95.47    0.14
 107 OMEGA       223 197  86   9 143 145    2.33   -0.03   34.12   34.20    0.78
 108 RHO+        213 197  86   9 146 147    2.07   -0.06   35.58   35.65    0.77
.......
 123 B-         -521 196  94  43 163 165  -25.44  -14.98   11.97   32.29    5.28
 .......
 136 B_S0        531 200 101  32 176 176   14.56  -33.17   -4.60   36.91    5.38
.......
 141 PI0         111   1 104   9   0   0    0.66    0.06   16.64   16.65    0.14
 142 RHO-       -213 197 104   9 185 186    1.47   -0.05   28.37   28.42    0.77

                       ---STRONG HADRON DECAYS---

   The unstable hadrons decay according to decay tables.  Remember that
   the pi0 was set stable in the initialization phase.  For heavy (b,c)
   quarks, partonic or direct hadronic decays may occur.  In this event
   the B- does a b -> u directly to rho0 e- nu_ebar. The B_s oscillates
   into a B_sbar which decays partonically to c cbar s sbar.

 IHEP    ID    IDPDG IST MO1 MO2 DA1 DA2    P-X     P-Y     P-Z   ENERGY   MASS 
 143 PI+         211   1 107   9   0   0    1.90   -0.02   24.27   24.35    0.14
 144 PI-        -211   1 107   9   0   0    0.32    0.00    6.77    6.78    0.14
 145 PI0         111   1 107   9   0   0    0.11   -0.02    3.07    3.08    0.14
 146 PI+         211   1 108   9   0   0    1.80    0.19   29.01   29.07    0.14
 147 PI0         111   1 108   9   0   0    0.27   -0.25    6.57    6.58    0.14
.......
 163 RHO0        113 198 123  43 194 195  -18.44   -9.34    6.79   21.77    0.77
 164 E-           11   1 123  43   0   0   -6.49   -5.26    5.27    9.88    0.00
 165 NU_EBAR     -12   1 123  43   0   0   -0.50   -0.38   -0.09    0.63    0.00
.......
 176 B_SBAR0    -531 199 136  32 207 208   14.56  -33.17   -4.60   36.91    5.38
 205 PI0         111   1 187   9   0   0    0.51   -0.06    1.56    1.65    0.14
 206 PI0         111   1 187   9   0   0    0.14    0.12    1.63    1.64    0.14

                       ---HEAVY FLAVOUR DECAYS--- 

 IHEP    ID    IDPDG IST MO1 MO2 DA1 DA2    P-X     P-Y     P-Z   ENERGY   MASS 
 207 BQRK          5 155 176 208 209 211   13.20  -30.08   -4.17   33.47    4.88
 208 SBAR         -3 125 176 211 212 211    1.35   -3.09   -0.43    3.43    0.50
 209 CQRK          4 123 207 210 213 210    2.30   -5.94   -0.61    6.59    1.55
 210 CBAR         -4 124 207 209 215 209    3.58   -8.37   -1.74    9.40    1.55
 211 SQRK          3 124 207 207 217 207    7.33  -15.77   -1.82   17.49    0.50
 212 SBAR         -3 160 208 221 223 221    1.35   -3.09   -0.43    3.43    0.50

   The B_sbar decay products hadronize to eta_c K+ K-:

                          ---CLUSTER DECAYS---    

 IHEP    ID    IDPDG IST MO1 MO2 DA1 DA2    P-X     P-Y     P-Z   ENERGY   MASS 
 224 ETA_C       441 199 222 213 227 229    4.87  -12.18   -2.12   13.62    2.98
 225 K-         -321   1 223 217   0   0    8.22  -17.67   -2.03   19.60    0.49
 226 K+          321   1 223 217   0   0    1.47   -3.31   -0.45    3.69    0.49

   The eta_c decays partonically to 3 gluons:

                   ---HEAVY FLAVOUR DECAYS--- 

 IHEP    ID    IDPDG IST MO1 MO2 DA1 DA2    P-X     P-Y     P-Z   ENERGY   MASS 
 227 GLUON        21 123 224 229 230 228    2.59   -4.45   -1.11    5.32    0.75
 228 GLUON        21 124 224 227 232 229    0.84   -4.20   -0.51    4.38    0.75
 229 GLUON        21 124 224 228 234 227    1.44   -3.53   -0.50    3.92    0.75

   Finally the 3 gluons hadronize to pi+ pi- 4 pi0:

                          ---CLUSTER DECAYS---    

 IHEP    ID    IDPDG IST MO1 MO2 DA1 DA2    P-X     P-Y     P-Z   ENERGY   MASS 
 245 PI-        -211   1 242 230   0   0    1.53   -3.79   -0.46    4.11    0.14
 246 PI0         111   1 242 230   0   0    0.06   -0.12   -0.07    0.20    0.14
 247 PI0         111   1 243 232   0   0    0.52   -0.39    0.03    0.66    0.14
 248 PI0         111   1 243 232   0   0    1.30   -3.27   -1.01    3.66    0.14
 249 PI0         111   1 244 234   0   0    1.02   -3.98   -0.74    4.18    0.14
 250 PI+         211   1 244 234   0   0    0.43   -0.64    0.14    0.80    0.14

   After the 1000 events requested  have been generated,  an analysis of
   the associated weight distribution and cross section is printed.